%% file: main.tex
\documentclass[aps,prx,reprint,showpacs,twocolumn,superscriptaddress,eqsecnum]{revtex4-2}

\usepackage{graphicx}
\usepackage[dvipsnames]{xcolor}
\usepackage{rotating}
\usepackage{amsmath,amssymb,graphics,amsthm,isomath}
\usepackage{amsfonts,dsfont,mathtools}
\usepackage{bbm}
\usepackage{bm}
\usepackage{float}
\usepackage{array}
\usepackage{appendix}
\newcolumntype{P}[1]{>{\centering\arraybackslash}p{#1}}
\usepackage{multirow}
\usepackage{physics}
\usepackage{braket}
\usepackage{adjustbox}
\usepackage{stmaryrd}
\usepackage{blkarray, bigstrut}
\usepackage{tikz-cd}
\usepackage{booktabs}
\usepackage{afterpage}
\usepackage{quantikz}
\usepackage{multirow}
\usepackage{colortbl}
\usepackage[colorlinks=true, urlcolor=blue, linkcolor=blue, citecolor=blue, hyperindex=true, linktocpage=true]{hyperref}
\usepackage[capitalise,compress]{cleveref}
\usepackage{enumitem}
\setlist[itemize]{label=$\bullet$}

\usepackage[normalem]{ulem}
\usepackage{comment}

\newtheorem*{theorem*}{Theorem}

\newcommand{\chelix}{C$_4$-Helix}

\begin{document}

\title{Experimental validation of a compact fault-tolerant architecture for trapped ions}

\author{Noah Berthusen}
\affiliation{ Quantinuum, Broomfield, CO 80021, USA}
\author{Ali Lavasani}
\affiliation{ Quantinuum, Broomfield, CO 80021, USA}
\author{Asmae Benhemou}
\affiliation{ Quantinuum, Partnership House, Carlisle Place, London, SW1P 1BX, UK}
\author{M.S. Allman}
\affiliation{ Quantinuum, Broomfield, CO 80021, USA}
\author{Joan Dreiling}
\affiliation{ Quantinuum, Broomfield, CO 80021, USA}
\author{Brian Estey}
\affiliation{ Quantinuum, Broomfield, CO 80021, USA}
\author{Cameron Foltz}
\affiliation{ Quantinuum, Broomfield, CO 80021, USA}
\author{Trent Jacobs}
\affiliation{ Quantinuum, Broomfield, CO 80021, USA}
\author{Michael Mills}
\affiliation{ Quantinuum, Broomfield, CO 80021, USA}
\author{Annie Jihyun Park}
\affiliation{ Quantinuum, Broomfield, CO 80021, USA}
\author{Adam P. Reed}
\affiliation{ Quantinuum, Broomfield, CO 80021, USA}
\author{David Hayes}
\affiliation{ Quantinuum, Broomfield, CO 80021, USA}
\author{Tzvetan S. Metodi}
\affiliation{ Quantinuum, Broomfield, CO 80021, USA}
\author{Andrew C. Potter}
\affiliation{ Quantinuum, Broomfield, CO 80021, USA}

\begin{abstract}
Quantum error correction (QEC) is beginning to enable logical operations that outperform their unencoded physical counterparts, but useful fault-tolerant computation will require more than low-error quantum memory. An effective architecture must orchestrate efficient logical encoding, low-overhead logical operations, and access to the non-Clifford resources required for universal computation. Here, we introduce and experimentally validate such an architecture based on the $[[20,2,6]]$ \chelix\ code, designed for the early fault-tolerant regime. Using Quantinuum Helios, a 98-qubit trapped-ion quantum processor, we experimentally demonstrate the principal components of this architecture: we perform repeated quantum error correction with an error of $4.6^{+6.2}_{-2.6}\times10^{-5}$ per logical qubit per QEC cycle. We benchmark the complete Clifford group on the two logical qubits of a single codeblock under active error correction, obtaining an error of $2.8^{+1.0}_{-1.6}\times 10^{-4}$ per two-qubit logical Clifford. We further demonstrate a fault-tolerant chain-map interface between \chelix\ and a distance-5 surface code, preparing a heterogeneous three-logical-qubit GHZ state with a fidelity lower bound of $99.925^{+0.068}_{-0.245}\%$. In each case, the encoded implementation outperforms its corresponding unencoded physical baseline without relying on postselection. Circuit-level simulations indicate that improvements in physical fidelity bring the same architecture into the $10^{-6}$--$10^{-8}$ logical-error regime targeted for early fault-tolerant computation. Together, these results establish \chelix\ as a hardware-validated fault-tolerant architecture rather than a bare quantum memory.
\end{abstract}


\maketitle


\section{Introduction}

Harnessing the full potential of quantum computers requires encoding many noisy physical qubits into a smaller number of logical qubits protected by quantum error correction (QEC)~\cite{shor1995}.
Encoding and performing fault-tolerant (FT) computation requires significant spatial (qubit number) and temporal (circuit complexity) overhead.
Above a physical error threshold, this additional overhead introduces more errors than are corrected, and only recently have demonstrations emerged in which QEC components have begun to break-even over their unencoded physical counterparts~\cite{anderson2021, Sivak_2023, Gupta2024, Hong_2024, bedalov2024, paetznick2024, google2024, Lacroix_2025, reichardt2025, dasu2025, dasu2026, Sivak2026}. 
Atomic platforms with arbitrary qubit connectivity offer a distinctive pathway to reduce these overheads by enabling access to long-range connected codes that provide much higher encoding rates than conventional planar topological codes, as well as access to fast logical operations through transversal and permutation gates.
While demanding applications like factoring and high-precision chemical modeling require extremely low logical error rates, the early-FT regime with modest LERs in the range of $10^{-6}$--$10^{-8}$ per logical gate would already unlock a variety of scientifically and commercially valuable computations beyond the reach of even the largest classical supercomputers.

Here, we introduce and experimentally validate a FT architecture suitable for this early-FT regime based on the \chelix\ code: a [[20,2,6]] concatenated symplectic double (CSD) code~\cite{berthusen2025simple} arising from concatenation of a [[10,2,3]] twisted Toric code, with the  [[4,2,2]] C$_4$ code~\cite{Vaidman_1996, Grassl_1997}. The \chelix\ code provides a $\sim 3.5\times$ reduction in spatial overhead compared to conventional rotated surface codes~\cite{bravyi1998quantum, Kitaev_2003} of comparable distance, achieving an encoding rate competitive with bivariate- and generalized-bicycle codes~\cite{Kovalev_2013, bravyi2023highthreshold}.
In contrast to many high-rate codes, in which logical computation is mediated predominantly through generalized lattice surgery~\cite{Cohen_2022, he2025extractors, zheng2025, baspin2025, cowtan2025}, the \chelix\ code permits efficient logical Clifford computation almost entirely through depth-1 transversal gates and automorphisms, with the latter requiring only physical single-qubit gates and qubit relabeling.
Beyond the code itself, we provide a complete blueprint for a fault-tolerant architecture built around the \chelix\ code, combining its native transversal and automorphism operations with a depth-4 fold-transversal~\cite{Breuckmann_2024} S-gate for complete Clifford control and chain-map CNOT interfaces~\cite{benhemou2026} to codes capable of magic-state production for universal computation.

\begin{figure*}
    \centering
    \includegraphics[width=0.8\linewidth]{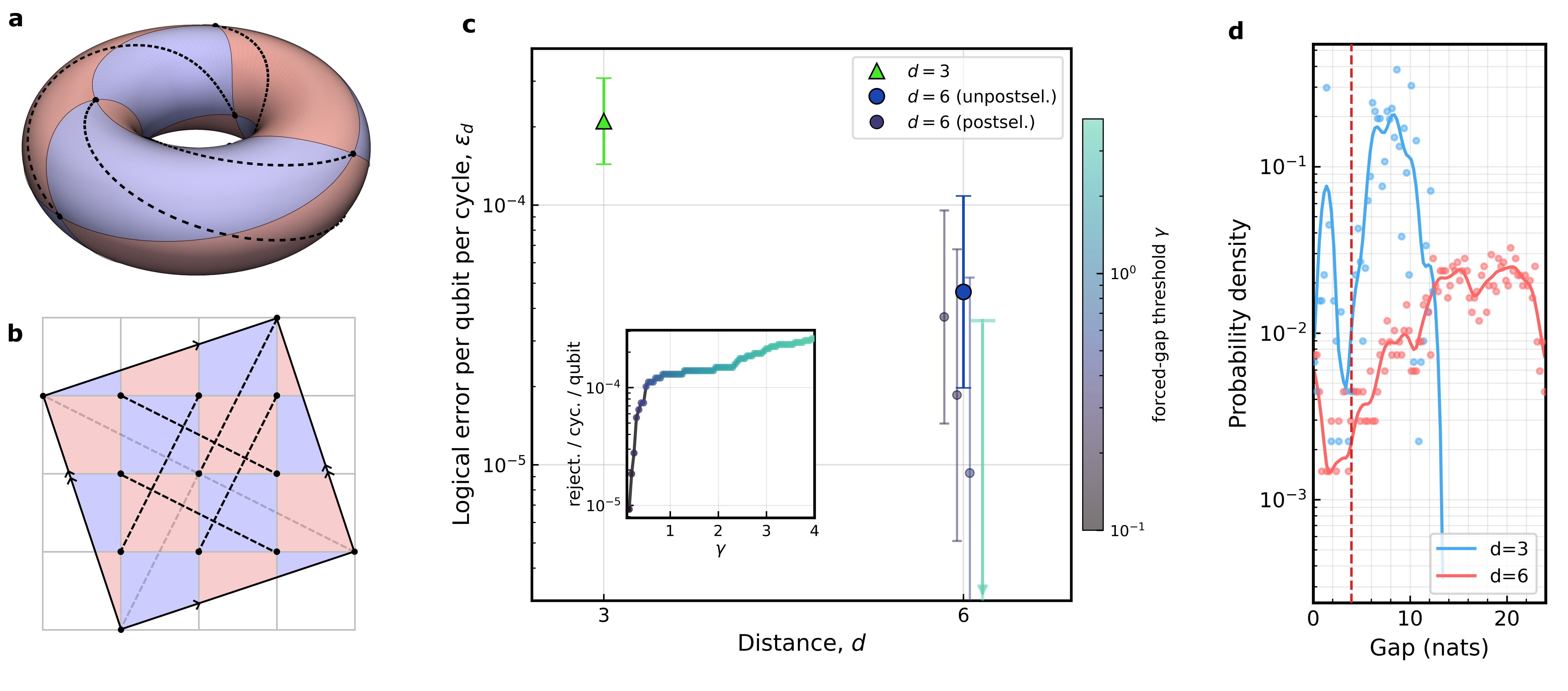}
    \caption{\textbf{\chelix\ code memory and distance scaling. a.} Schematic of the $[[10,2,3]]$ and $[[20,2,6]]$ \chelix\ codes having been mapped onto a torus. The underlying lattice corresponds to the plaquettes of the $[[10,2,3]]$ twisted Toric code. Dashed lines represent the pairing taken when performing the $C_4$ concatenation for the $C_4$-Helix code. \textbf{b.} Same schematic of the Helix code sequence, but on a flattened torus. \textbf{c.} Logical error rate per LQ, per cycle $\varepsilon_d$ as a function of the code distance. We report results for the $[[10,2,3]]$, and $[[20,2,6]]$ \chelix code with/out forced-gap postselection. All uncertainties reported through this work denote 95\% Wilson confidence intervals unless otherwise noted. \textbf{d.} Probability distribution of the Frontier decoder's confidence gap between logical classes, in nats. Specifying a postselection parameter $\gamma$ corresponds to discarding shots that lie to the left of a vertical line at $x=\gamma$. Plotted as the red, vertical, dashed line is $\gamma = 3.9$, the smallest gap that postselects out every logical failure in the \chelix\ code.
    }
    \label{fig:memory_main}
\end{figure*}

We then implement and benchmark the essential components of this early-FT architecture on Quantinuum Helios: a 98-qubit trapped-ion quantum processor (QPU)~\cite{Ransford2026}. 
First, we benchmark repeated QEC in a memory setting and observe logical-error suppression under $C_4$ concatenation, reducing the per LQ, per QEC cycle error from $2.1^{+1.0}_{-0.7}\times10^{-4}$ for the $[[10,2,3]]$ code to $4.6^{+6.2}_{-2.6}\times10^{-5}$ for the $[[20,2,6]]$ \chelix\ code.
Second, we test the computational model directly by performing two-qubit randomized benchmarking (TQRB)~\cite{magesan2011, combes2017} of the complete logical Clifford group while interleaving active adaptive syndrome extraction~\cite{Berthusen_2025}, achieving an error per Clifford of $\varepsilon_C = 2.8^{+1.0}_{-1.6}\times 10^{-4}$, well below the physical unencoded TQRB result: $1.2\times 10^{-3}$.
Third, we test the heterogeneous-code interface required for magic state injection by using a chain-map CNOT to prepare a three-logical-qubit GHZ state spanning \chelix\ and a distance-5 rotated surface code, obtaining a logical GHZ-fidelity lower bound of $99.93\pm0.04\%$, which again improves upon the physical result of $99.54 \pm 0.04\%$.

Finally, circuit-level simulations show that moderate improvements in physical gate fidelity, already achieved in trapped-ion testbeds~\cite{hughes2025}, can reduce logical error rates by orders of magnitude and bring the same $[[20,2,6]]$ code into the $10^{-6}$--$10^{-8}$ regime.
Together, these results establish \chelix\ as a hardware-validated architecture for early-FT quantum computation, combining protected memory, complete low-overhead Clifford control, and a FT interface to non-Clifford resources.

\section{Logical Gadgets}

\begin{figure*}
    \centering
    \includegraphics[width=\linewidth]{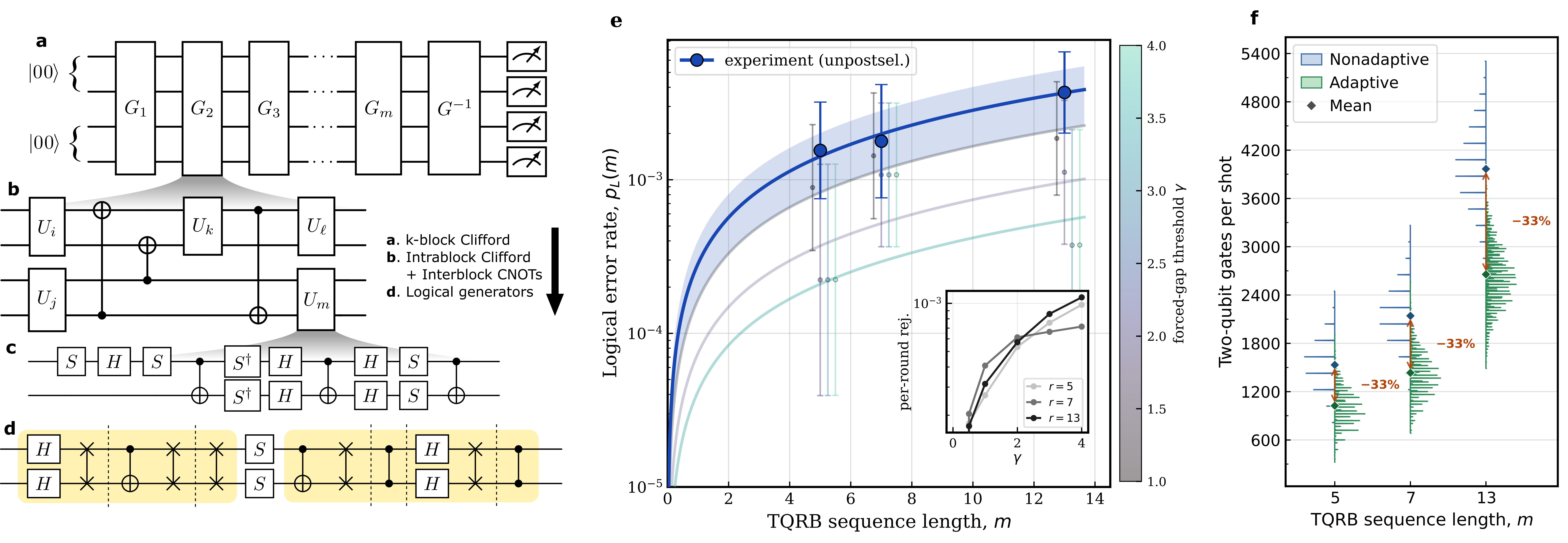}
    \caption{\textbf{Logical randomized Clifford benchmarking. a.} A RB experiment on four qubits, where $G_i \in \mathcal{C}_4$ and $G_1 G_2...G_mG^{-1} = I$. \textbf{b.} Each four-qubit Clifford is broken up into two-qubit Cliffords, $U \in \mathcal{C}_2$, on each \chelix\ codeblock and interblock CNOT gates. \textbf{c.} A two-qubit Clifford is decomposed into single-qubit and CNOT gates. \textbf{d.} The decomposed two-qubit Clifford is expressed in the logical gateset of the \chelix\ code: $\{ \overline{H}_0\overline{H}_1\overline{SWAP}_{0,1}$, $\overline{CNOT}_{0,1}\overline{SWAP}_{0,1}$, $\overline{SWAP}_{0,1}$, $\overline{CZ}_{0,1}$, $\overline{S}_0 \overline{S}_1 \}$. \textbf{e.} Logical circuit error rate $p_L(m)$ as a function of TQRB sequence length. Logical error rates can again be reduced via forced-gap postselection for a fixed per-round rejection rate (see inset). \textbf{f.} Distribution of physical two-qubit gates per shot for a nonadaptive TQRB protocol versus the adaptive protocol that was run on Helios. The real-time circuit pruning done by ASE yields a reduction in the two-qubit gate count by $33\%$, while also reducing the wall-clock shot time by $23\%$.}
    \label{fig:clifford_bench_main}
\end{figure*}

\subsection{Error-corrected memory}

First, we benchmark the memory capabilities of the \chelix\ code and demonstrate logical-error suppression under $C_4$ concatenation, increasing the code distance from [[10,2,3]] to [[20,2,6]]. 
The memory experiments proceed as follows: we prepare the codes in the $\ket{++}$ logical state by projectively measuring the $Z$-type stabilizer generators. We then perform $r=20$ rounds of syndrome extraction in which we have paid special attention to mitigating leakage, a dominant error source on Helios. When a qubit leaks out of the computational subspace, any subsequent gates involving the leaked qubit can fail to enact their intended operation, producing persistent and temporally correlated faults. To mitigate leakage, we leverage a hardware-based leakage repump capability on Helios~\cite{lavasani2026} as well as circuit-level leakage reduction units (LRUs)~\cite{paetznick2024}; see Appendix~\ref{apx:syndrome_extraction}. 
Finally, we destructively measure the data qubits in the $X$-basis, from which we can construct a final round of $X$-type stabilizer generators as well as the $X$-type logical operators. 
We decode offline over the entire circuit volume using the Frontier decoder~\cite{leverrier2026}. The block logical error rate for an $r$-round memory experiment, $p_L(r)$, is the fraction of shots that have a logical error after decoding. We extract the block logical error per round for each effective distance, $\varepsilon_{d,\text{block}}$, according to the equation: $\varepsilon_{d,\text{block}} = 1 - (1 - p_L(r))^{1/r}$. Throughout, we report the corresponding per LQ error rate normalized by the number of encoded logical qubits $\varepsilon_d = \varepsilon_{d,\text{block}} / k$.

The results are shown in Fig.~\ref{fig:memory_main}(c). For the $[[10,2,3]]$ code, we obtain a per LQ, per round error of $\varepsilon_3 = 2.1^{+1.0}_{-0.7}\times10^{-4}$. 
For the $[[20,2,6]]$ \chelix\ code, we obtain $\varepsilon_6 = 4.6^{+6.2}_{-2.6}\times10^{-5}$.
All uncertainties are reported as 95\% Wilson confidence intervals. This $4.52\times$ reduction in logical error rate per round from $\varepsilon_3$ to $\varepsilon_6$ demonstrates logical-error suppression under $C_4$-concatenation at the operating point of Helios. 
We only sample the $X$-basis, and as such are unable to fully characterize the logical Pauli error channel; however, since the dominant memory error on Helios is pure dephasing ($Z$), the $X$-basis experiment is expected to provide a conservative proxy for the worse-performing memory basis.

A distance-6 code can correct all double faults and detect but not necessarily correct all triple faults.
We exploit this additional information using forced-gap postselection. Specifically, the Frontier decoder~\cite{leverrier2026} provides a confidence gap between competing logical classes, allowing shots with low-confidence corrections to be rejected; see Appendix~\ref{apx:decoder}. Related confidence-based postselection procedures have been proposed for other decoders~\cite{gidney2023, lee2026, xie2026, wills2026}. Applying forced-gap postselection to the Helios memory experiment allows us to dramatically reduce the logical error rate at a relatively minor discard rate: indeed, by only postselecting on 0.4\% of shots ($1\times 10^{-4}$ amortized per LQ, per round), we obtain a logical error rate of $\varepsilon_{6,PS} = 9.3^{+43}_{-7.6}\times 10^{-6}$. Although the absolute rejection rate is small, it exceeds the $\sim$0.2\% logical-failure probability of the unpostselected 20-round experiment, illustrating the tradeoff between logical accuracy and sampling overhead.

\subsection{Clifford randomized benchmarking}

A central feature of the \chelix\ architecture is that encoding multiple logical qubits does not require correspondingly expensive logical computation. By construction, CSD codes have a variety of CNOT, Hadamard, and CZ-type logical gates all implementable with only physical single-qubit gates and qubit relabeling. These SWAP-transversal, or automorphism, gates provide the ability to do some logical circuits essentially for free, as permutations are facilitated by ion-transport and qubit-relabeling in software. The automorphism gates of the \chelix\ code, when supplemented with a depth-4, fold-transversal~\cite{Breuckmann_2024}, global phase gate ($\overline{S}_0\overline{S}_1$), generate the full Clifford group on its two logical qubits. Moreover, because the two-qubit Clifford symplectic group is small, we can brute-force enumerate an $\overline{S}_0\overline{S}_1$-optimal decomposition of every Clifford and construct a lookup-table compiler. 
Together with transversal CNOTs or interblock chain maps~\cite{benhemou2026}, these operations extend to Clifford computation across multiple \chelix\ codeblocks.

We experimentally test this computational model using two-qubit randomized benchmarking (TQRB) on the two logical qubits of a single \chelix\ codeblock. As illustrated in Fig.~\ref{fig:clifford_bench_main}(a), each benchmark consists of a sequence of $m$ random Clifford operators $G_i \in \mathcal{C}_2$, followed by an inverse Clifford $G^{-1}$ such that the ideal logical action of the complete sequence is the identity. Previous experimental demonstrations of logical randomized benchmarking have used $k=1$ color-code blocks~\cite{mayer2024, Lacroix_2025}, whose transversal Clifford gates make such benchmarks particularly convenient. The similarly efficient logical gates of the \chelix\ code allow us to extend this approach to the complete two-logical-qubit Clifford group encoded within a single $k=2$ codeblock.



For each sampled Clifford $G_i$, we query the lookup-table compiler to obtain a sequence of native \chelix\ logical gates, as illustrated in Fig.~\ref{fig:clifford_bench_main}(c)--(d). The resulting sequence may contain up to three applications of the non-automorphism gate $\overline{S}_0\overline{S}_1$, each followed by a round of adaptive syndrome extraction (ASE)~\cite{Berthusen_2025}. The full length-$m$ Clifford sequence is generated and executed at runtime using Guppy~\cite{koch2025}, yielding an independent random sequence for every shot, while we track its accumulated symplectic action to construct the final inverse Clifford that ideally returns the logical state to $\ket{\overline{00}}$, up to Pauli corrections. Because the intervening logical gates transform both the stabilizer generators and logical observables, the corresponding detector structure depends on the particular Clifford sequence executed and is updated accordingly; details of this sign tracking are given in Appendix~\ref{apx:simulations}. We then decode the complete circuit volume using the Frontier decoder and count a shot as a logical error if the expected initial logical state is not recovered.

We perform TQRB at sequence lengths $m=\{5,7,13\}$ (including the final corrective Clifford), with the longest sequences requiring up to 27 rounds of adaptive syndrome extraction. Fitting the observed survival probabilities to 
$
  p_L(m) = \tfrac{1}{2}\left[\,1 - (1-2\varepsilon_C)^{\,m}\,\right]$,
we extract a logical error rate per Clifford of $\varepsilon_C = 2.8^{+1.0}_{-1.6}\times 10^{-4}$. Asymmetric 95\% confidence intervals are obtained via standard parametric bootstrap. This is a $4.3^{+5.9}_{-1.1} \times$ improvement over Helios' physical TQRB error per Clifford of $\approx 1.2\times 10^{-3}$~\cite{Ransford2026}, achieved without postselection. The performance is partially enabled by ASE: as shown in Fig.~\ref{fig:clifford_bench_main}(f), ASE reduces the number of physical two-qubit gates per $\overline{S}_0 \overline{S}_1$ gate by $\sim 33\%$. This pruning also shortens the physical run time, lowering the mean per-shot wall-clock duration by $\sim 23\%$. Two-qubit gates and idling significantly contribute to the physical error budget, and so these reductions translate into a lower logical error rate. While only the adaptive scheme was run on hardware, simulations (Fig.~\ref{fig:logic_sims_adapt}) indicate that the non-adaptive protocol would have performed worse under the same conditions, a gap that is expected to widen as the physical error rate improves.

The same compilation strategy extends naturally to multiple \chelix\ codeblocks. An arbitrary $m$-block Clifford $G \in \mathcal{C}_{2m}$ onto the \chelix\ gateset can be decomposed into two-qubit Cliffords $U_i \in \mathcal{C}_2$ fully supported on each \chelix\ codeblock and CNOT (CZ) gates between codeblocks~\cite{Aaronson_2004, maslov2017}. Either these inter-block CNOT gates are transversal, or targeted logical CNOT gates are required, Fig.~\ref{fig:cnot-circuit}. Then, each $U_i$ is decomposed into the native logical generators by means of the lookup-table compiler. This entire procedure is illustrated in Fig.~\ref{fig:clifford_bench_main}(a)-(d).
We note that the uniformly random two-qubit Cliffords used in this benchmark constitute a worst case workload for a single \chelix\ codeblock. The per-Clifford logical fidelities reported here should therefore be regarded as a conservative lower bound. The individual Clifford operations arising in practical quantum algorithms, whose structure can be compiled to favor cheap automorphism gates, are implemented at substantially higher fidelity and lower cost~\cite{LeBlond2026InPrep}.


\subsection{Inter-code GHZ state}
\label{sec:ghz}

\begin{figure}[t]
\centering

\includegraphics[width=0.9\linewidth]{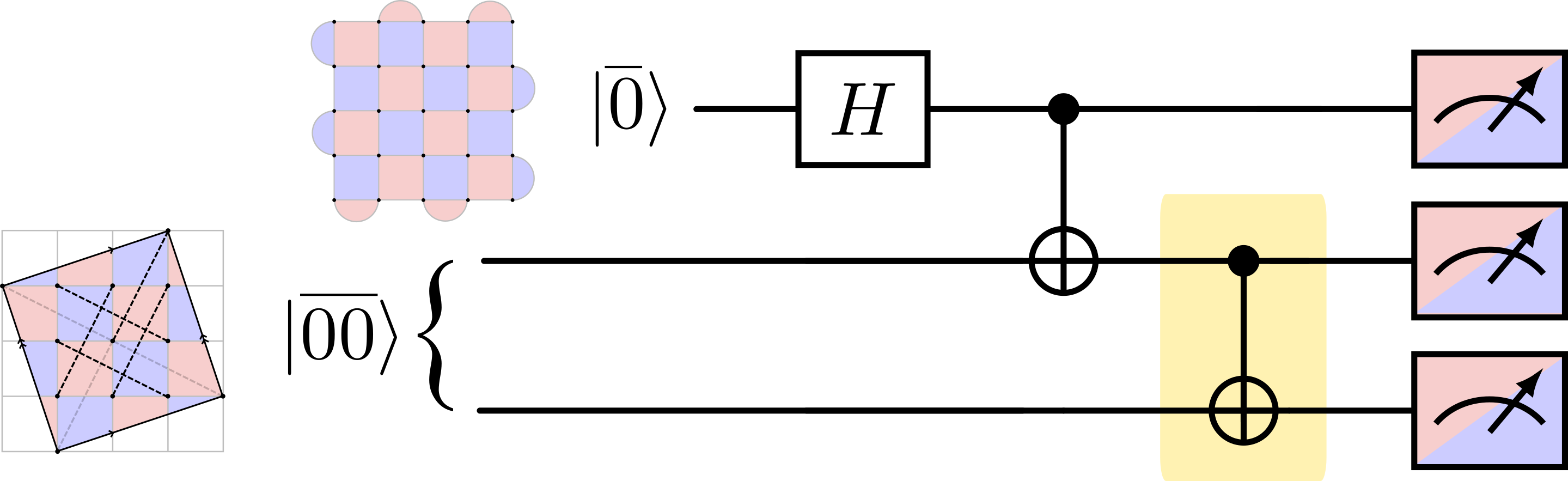}

\vspace{12pt}

\begin{tabular}{lcc}
\toprule
GHZ experiment & $\bar{z}$ mismatch (\%) & $\bar{x}$ mismatch (\%) \\
\midrule
Physical & $0.209^{+0.056}_{-0.047}$ & $0.253^{+0.061}_{-0.052}$ \\
Logical  & $0.025^{+0.114}_{-0.024}$ & $0.050^{+0.130}_{-0.044}$ \\
\bottomrule
\end{tabular}

\caption{
\textbf{Inter-code GHZ experiment schematic and results.} For each experiment we report $\bar{z} = \Tr(P_Z^-\rho)$ and $\bar{x} = \Tr(P_X^-\rho)$, the fraction of shots violating the $Z$-basis agreement and $X$-basis parity conditions,
respectively. The physical baseline comprises 32 three-qubit GHZ states prepared in parallel and sampled over $N=32{,}000$ groups per basis; the logical experiment uses the three-qubit logical GHZ state sampled over $N=4{,}000$ shots per basis. Uncertainties are $95\%$ Clopper--Pearson intervals.}
\label{fig:ghz_results}
\end{figure}

Efficient Clifford operations alone are insufficient for universal fault-tolerant computation; the architecture must also provide access to non-Clifford resource states. Rather than requiring such states to be prepared directly within the \chelix\ code, we consider an architecture in which magic states are prepared in a code optimized for that task and subsequently transferred into \chelix. Several approaches can mediate computation across code boundaries. Code switching temporarily transforms a logical state from one code into another~\cite{pogorelov2024, daguerre2025, Bluvstein2026}, while homomorphic CNOT gates directly couple logical qubits encoded in related CSS codes~\cite{huang2023}. Chain-map CNOTs generalize this latter approach to distinct CSS codes, enabling fault-tolerant entangling operations between codes with different block sizes and stabilizer structure~\cite{benhemou2026}. This provides a natural interface for a heterogeneous architecture in which \chelix\ serves as the computational code while non-Clifford resource states are prepared elsewhere.

Using an optimization-based method for finding chain maps~\cite{benhemou2026}, we obtain a fault-tolerant, depth-2 logical CNOT between a $[[25,1,5]]$ rotated surface code and the first logical qubit of the $[[20,2,6]]$ \chelix\ code. This operation provides the inter-code primitive required by our proposed state-injection scheme. Here, rather than performing magic-state injection directly, we experimentally benchmark the interface by using the chain-map CNOT to prepare a three-logical-qubit GHZ state spanning the two code families.

As illustrated in Fig.~\ref{fig:ghz_results}, we prepare a three-logical-qubit GHZ state between the surface-code logical qubit and the two logical qubits of the \chelix\ code. After preparing the rotated surface code and \chelix\ code in the $Z$-basis, we apply a SWAP-transversal logical Hadamard gate~\cite{chen2026} to the surface code followed by the chain map implementing the first CNOT gate. The second CNOT between the two \chelix\ logical qubits is implemented as an automorphism and therefore requires only qubit relabeling. Finally, we destructively measure out both codes in either the $X$- or $Z$-basis and construct the joint logical operators $\{\overline{Z}_0\overline{Z}_1, \overline{Z}_1 \overline{Z}_2, \overline{X}_0 \overline{X}_1\overline{X}_2\}$. Prior to checking for valid logical measurement outcomes, we apply a correction by decoding over the entire circuit volume. As an unencoded baseline, we prepare 32 three-qubit physical GHZ states in parallel across 96 of the 98 qubits on Helios and measure them in the same bases.

To characterize the resulting state, we follow the approach from Ref.~\cite{Hong_2024}. An $n$-qubit GHZ state is stabilized by
$\{Z_iZ_{i+1}\}_{i=0}^{n-2}$ together with
$X_0X_1\cdots X_{n-1}$. Accordingly, a perfect GHZ state produces equal outcomes in the $Z$ basis and even parity in the $X$ basis. Defining $\bar{z}=\Tr(P_Z^-\rho)$ and $\bar{x}=\Tr(P_X^-\rho)$, where $P^{-}_{Z/X}$ is the projector into $-1$ eigenspace in a given basis, as the probabilities of violating these conditions, the GHZ-state fidelity is bounded by
\begin{equation}
    1-\Tr(P_Z^- \rho) - \Tr(P_X^- \rho) \le F_{\rm{GHZ}}(\rho) \le 1- \Tr(P_Z^- \rho).
\end{equation}

The results are shown in Fig.~\ref{fig:ghz_results}, from which we obtain the bounds
\begin{align}
    99.537^{+0.099}_{-0.118}\% &\le F_\text{GHZ}(\rho) \le 99.791^{+0.047}_{-0.056}\% \\
    99.925^{+0.068}_{-0.245}\% &\le F_\text{GHZ}(\overline{\rho}) \le 99.975^{+0.024}_{-0.114}\%\end{align}
where $\rho$ and $\overline{\rho}$ denote the physical (unencoded) and logical GHZ states, respectively. The logical-state lower bound exceeds the physical baseline by $\Delta F\approx0.39\%$, corresponding to a $6.7\sigma$ separation, with non-overlapping 95\% confidence intervals. Because this fidelity bound includes state preparation, logical operations, the surface-code--\chelix\ chain map, and readout, it bounds the aggregate error of the complete logical sequence rather than the chain map in isolation. Nevertheless, the result rules out any large degradation associated with the heterogeneous-code interface and experimentally validates the entangling primitive required by the architecture's proposed route to magic-state injection.

\section{Discussion and outlook}
\label{sec:discussion}

In this work, we have experimentally validated the principal components of a fault-tolerant architecture based on the $[[20,2,6]]$ \chelix\ code: repeated error correction with logical-error suppression under $C_4$ concatenation, randomized benchmarking of the complete Clifford group on two logical qubits encoded within a single $k=2$ codeblock, and entangling operations between \chelix\ and a distinct code family through a chain-map construction. In each experiment, the encoded implementation outperformed the corresponding unencoded physical baseline on Helios, without postselection. A central feature of this architecture is that complete Clifford control is obtained with low logical-gate overhead: the Clifford generators are implemented using automorphisms, transversal operations, and a single depth-4 fold-transversal phase-gate construction, avoiding the ancilla resources and serialization associated with generalized lattice-surgery approaches to computation on many high-rate codes.

Universal computation additionally requires access to non-Clifford resource states. The chain-map interface demonstrated here provides a mechanism for supplying these resources from a code optimized for their preparation, rather than requiring magic-state production within \chelix\ itself. In particular, a cultivated magic state could be prepared in a surface- or color-code block~\cite{gidney2024} and coupled to \chelix\ using the same type of heterogeneous-code CNOT demonstrated in Section~\ref{sec:ghz}. Other source codes, including codes with transversal non-Clifford gates such as triorthogonal codes~\cite{Bravyi_2012}, can be incorporated using chain maps obtained through the same optimization procedure~\cite{benhemou2026}. Alternatively, non-Clifford resources may be prepared directly within \chelix\ through zero-level distillation~\cite{Itogawa_2025} of a logical $\ket{\overline{CZ}}$ state~\cite{Gupta2024}; see Appendix~\ref{apx:non-clifford}. The present experiment does not implement a universal Clifford+$T$ computation, but establishes the Clifford computational substrate and heterogeneous-code interface needed to construct one.

The architecture can also be extended to larger code distances while retaining much of the logical-gate structure of \chelix. In the CSD construction, the inner $C_4$ $[[4,2,2]]$ code can be replaced by higher-distance $k=2$ codes possessing the required $X/Z$ duality and SWAP-transversal two-qubit operations. Concatenating the $[[10,2,3]]$ Helix code with, for example, the $[[12,2,4]]$ Carbon code~\cite{paetznick2024} or with \chelix\ itself produces the $[[60,2,12]]$ Carbon-Helix and $[[100,2,18]]$ Double-Helix codes, respectively, while preserving much of the transversal and automorphism-based Clifford structure. We describe these higher-distance constructions and their logical compilation in Appendix~\ref{apx:big_$C_4$-Helix}; a detailed circuit-level assessment of their performance is left to future work.

Together, these results establish \chelix\ as a hardware-validated fault-tolerant architecture rather than solely a quantum memory. At the physical error rates demonstrated here, the \chelix\ experiments achieve logical error rates at the $\sim10^{-4}$ level per QEC cycle and logical Clifford. Circuit-level simulations indicate that improvements in physical infidelity to the $10^{-4}$ regime drive the same $[[20,2,6]]$ code into the $10^{-6}$--$10^{-8}$ logical-error regime targeted for early fault-tolerant computation, with still lower error rates accessible through modest decoder-based postselection; see Appendix~\ref{apx:simulations}. The higher-distance CSD constructions described here provide a complementary route to lower logical error rates while retaining much of the logical-gate and compilation structure of \chelix. More broadly, the results show that compact, $k>1$, quantum codes can be designed around the requirements of computation from the outset, rather than treating logical operations as an additional layer built on top of memory protection.


\section*{Acknowledgements}
We acknowledge the entire Quantinuum team for innumerable contributions towards the design, construction, and operation of the Helios quantum computer, and we acknowledge Honeywell for fabricating the trap used in this experiment. We thank Tyler LeBlond and Shival Dasu for insightful conversations.


\section*{Additional information}
Correspondence and requests for materials should be addressed to Noah Berthusen (\href{mailto:noah.berthusen@quantinuum.com}{noah.berthusen@quantinuum.com}) and Andrew C. Potter (\href{mailto:andrew.potter@quantinuum.com}{andrew.potter@quantinuum.com}).

\bibliography{bibliography}

\clearpage
\appendix
\input{appendix}

\end{document}

%% file: appendix.tex
\section{Details of the $[[20,2,6]]$ $C_4$-Helix code}
\label{apx:code_construction}

Here we provide additional details on the construction of the $[[20,2,6]]$ $C_4$-Helix code and its logical gadgets. We start the CSD construction~\cite{berthusen2025simple} with the well-known $[[5,1,3]]$ non-CSS code~\cite{laflamme1996} with symplectic parity check matrix:
\begin{align}
H_{[[5,1,3]]} =  \left(\begin{array}{c|c}
1 0 0 1 0 & 0 1 1 0 0 \\
0 1 0 0 1 & 0 0 1 1 0 \\
1 0 1 0 0 & 0 0 0 1 1 \\
0 1 0 1 0 & 1 0 0 0 1 
\end{array} \right).
\label{eq:513pcm}
\end{align}
We then obtain the $[[10,2,3]]$ symplectic double code~\cite{burton2024} by considering Eq.~\eqref{eq:513pcm} as $H = (H_X|H_Z)$ and constructing the following parity check matrix
\begin{equation}
\mathfrak{D}(H) = \begin{pmatrix}
\begin{array}{cc|cc}
H_X & H_Z & 0 & 0 \\
0 & 0 & H_Z & H_X \\
\end{array}
\end{pmatrix}.
\label{eq:pcm_double}
\end{equation}
Explicitly, the stabilizer generators and logical operators for the $[[10,2,3]]$ code are shown in the left panel of Table~\ref{tab:codes}, including the two redundant stabilizers resulting from the redundant stabilizer $ZZXIX$ in the $[[5,1,3]]$ code. The stabilizers of the $[[10,2,3]]$ code can be laid out to be planar on a twisted torus, as shown in Fig.~\ref{fig:memory_main}(b). It is for this reason that the $[[10,2,3]]$ code is called a twisted, or cyclic~\cite{Sarkar_2024} Toric code. Measuring the stabilizer generators of the twisted Toric code can be done using the standard ``N-Z" schedule for surface codes~\cite{Tomita_2014}. Projective state preparation must be done with a different stabilizer measurement schedule in order to be fault-tolerant.

The symplectic double construction gives the $[[10,2,3]]$ code a ZX-duality~\cite{Breuckmann_2024}, $\tau: \{i, i+5\}_{0,...,9}$. It is along this symmetry that we concatenate with the $[[4,2,2]]$, or $C_4$ code~\cite{Vaidman_1996, Grassl_1997} ,with logical operators defined to be $XIIX$, $IIZZ$, $IIXX$, $ZIIZ$. At a high level, the concatenation procedure can be thought of as replacing each physical qubit of the $[[10,2,3]]$ code with a logical qubit from a $C_4$ codeblock. Physical qubits are assigned to codeblocks according to $\tau$, e.g., physical qubits 0 and 5 are placed in the same $C_4$ codeblock. The assignment of logicals within a $C_4$ codeblock can be arbitrary, but we default to treating the lower indexed physical qubit as the 0th logical qubit. The resulting concatenated code then has two weight-4 generators coming from each of the five $C_4$ codeblocks and eight weight-8 generators obtained by including all $[[10,2,3]]$ stabilizer generators but replacing each single-qubit Pauli operator with the corresponding $C_4$ logical operator, e.g., $X_3$ is replaced with $\overline{X}_0$ on the fourth $C_4$ block---$X_{12}X_{15}$. We have 18 independent stabilizer generators, hence yielding a $[[20,2,6]]$ code. The $[[20,2,6]]$ code is self-dual with $H_X = H_Z$.

The exact stabilizer generators (including two that are redundant) are displayed in the right panel of Table~\ref{tab:codes}. Note that the concatenated stabilizers presented in Table~\ref{tab:codes} for the $[[20,2,6]]$ code are slightly different than what would immediately arise from the CSD construction. Instead, we use a representation where some concatenated generators have been multiplied by a $C_4$ stabilizer generator to facilitate more convenient syndrome extraction; see Section~\ref{apx:syndrome_extraction}.


\begin{table*}

\begin{minipage}{0.33\textwidth}
\raggedleft
\begin{tabular}{cccccccccccc}
0 & 1 & 2 & 3 & 4 & 5 & 6 & 7 & 8 & 9 & \\
\hline
$X$& $\cdot$& $\cdot$& $X$& $\cdot$& $\cdot$& $X$& $X$& $\cdot$& $\cdot$&\\
$\cdot$& $X$& $\cdot$& $\cdot$& $X$& $\cdot$& $\cdot$& $X$& $X$& $\cdot$&\\
$X$& $\cdot$& $X$& $\cdot$& $\cdot$& $\cdot$& $\cdot$& $\cdot$& $X$& $X$&\\
$\cdot$& $X$& $\cdot$& $X$& $\cdot$& $X$& $\cdot$& $\cdot$& $\cdot$& $X$&\\
$\cdot$& $\cdot$& $X$& $\cdot$& $X$& $X$& $X$& $\cdot$& $\cdot$& $\cdot$&\\
\hline
$\cdot$& $Z$& $Z$& $\cdot$& $\cdot$& $Z$& $\cdot$& $\cdot$& $Z$& $\cdot$&\\
$\cdot$& $\cdot$& $Z$& $Z$& $\cdot$& $\cdot$& $Z$& $\cdot$& $\cdot$& $Z$&\\
$\cdot$& $\cdot$& $\cdot$& $Z$& $Z$& $Z$& $\cdot$& $Z$& $\cdot$& $\cdot$&\\
$Z$& $\cdot$& $\cdot$& $\cdot$& $Z$& $\cdot$& $Z$& $\cdot$& $Z$& $\cdot$&\\
$Z$& $Z$& $\cdot$& $\cdot$& $\cdot$& $\cdot$& $\cdot$& $Z$& $\cdot$& $Z$&\\
\hline
$\cdot$& $\cdot$& $X$& $X$& $\cdot$& $X$& $\cdot$& $\cdot$& $\cdot$& $\cdot$& $=X_0$\\
$X$& $\cdot$& $\cdot$& $\cdot$& $\cdot$& $\cdot$& $X$& $\cdot$& $\cdot$& $X$& $=X_1$\\
$\cdot$& $Z$& $\cdot$& $\cdot$& $Z$& $Z$& $\cdot$& $\cdot$& $\cdot$& $\cdot$& $=Z_0$\\
$Z$& $\cdot$& $\cdot$& $\cdot$& $\cdot$& $\cdot$& $\cdot$& $Z$& $Z$& $\cdot$& $=Z_1$\\
\end{tabular}
\end{minipage}
\hfill
\begin{minipage}{0.6\textwidth}
\raggedright
\begin{tabular}{cccccccccccccccccccccc}
0 & 1 & 2 & 3 & 4 & 5 & 6 & 7 & 8 & 9 & 10 & 11 & 12 & 13 & 14 & 15 & 16 & 17 & 18 & 19 & \\
\hline
$X$& $X$& $X$& $X$& $\cdot$& $\cdot$& $\cdot$& $\cdot$& $\cdot$& $\cdot$& $\cdot$& $\cdot$& $\cdot$& $\cdot$& $\cdot$& $\cdot$& $\cdot$& $\cdot$& $\cdot$& $\cdot$&\\
$\cdot$& $\cdot$& $\cdot$& $\cdot$& $X$& $X$& $X$& $X$& $\cdot$& $\cdot$& $\cdot$& $\cdot$& $\cdot$& $\cdot$& $\cdot$& $\cdot$& $\cdot$& $\cdot$& $\cdot$& $\cdot$&\\
$\cdot$& $\cdot$& $\cdot$& $\cdot$& $\cdot$& $\cdot$& $\cdot$& $\cdot$& $X$& $X$& $X$& $X$& $\cdot$& $\cdot$& $\cdot$& $\cdot$& $\cdot$& $\cdot$& $\cdot$& $\cdot$&\\
$\cdot$& $\cdot$& $\cdot$& $\cdot$& $\cdot$& $\cdot$& $\cdot$& $\cdot$& $\cdot$& $\cdot$& $\cdot$& $\cdot$& $X$& $X$& $X$& $X$& $\cdot$& $\cdot$& $\cdot$& $\cdot$&\\
$\cdot$& $\cdot$& $\cdot$& $\cdot$& $\cdot$& $\cdot$& $\cdot$& $\cdot$& $\cdot$& $\cdot$& $\cdot$& $\cdot$& $\cdot$& $\cdot$& $\cdot$& $\cdot$& $X$& $X$& $X$& $X$&\\
$X$& $\cdot$& $\cdot$& $X$& $\cdot$& $\cdot$& $X$& $X$& $\cdot$& $\cdot$& $X$& $X$& $X$& $\cdot$& $\cdot$& $X$& $\cdot$& $\cdot$& $\cdot$& $\cdot$&\\
$\cdot$& $\cdot$& $\cdot$& $\cdot$& $X$& $\cdot$& $\cdot$& $X$& $X$& $X$& $\cdot$& $\cdot$& $\cdot$& $\cdot$& $X$& $X$& $X$& $\cdot$& $\cdot$& $X$&\\
$\cdot$& $X$& $X$& $\cdot$& $\cdot$& $\cdot$& $\cdot$& $\cdot$& $X$& $\cdot$& $\cdot$& $X$& $X$& $X$& $\cdot$& $\cdot$& $\cdot$& $\cdot$& $X$& $X$&\\
$\cdot$& $\cdot$& $X$& $X$& $\cdot$& $X$& $X$& $\cdot$& $\cdot$& $\cdot$& $\cdot$& $\cdot$& $\cdot$& $X$& $X$& $\cdot$& $X$& $X$& $\cdot$& $\cdot$&\\
$X$& $X$& $\cdot$& $\cdot$& $X$& $X$& $\cdot$& $\cdot$& $\cdot$& $X$& $X$& $\cdot$& $\cdot$& $\cdot$& $\cdot$& $\cdot$& $\cdot$& $X$& $X$& $\cdot$&\\
\hline
$Z$& $Z$& $Z$& $Z$& $\cdot$& $\cdot$& $\cdot$& $\cdot$& $\cdot$& $\cdot$& $\cdot$& $\cdot$& $\cdot$& $\cdot$& $\cdot$& $\cdot$& $\cdot$& $\cdot$& $\cdot$& $\cdot$&\\
$\cdot$& $\cdot$& $\cdot$& $\cdot$& $Z$& $Z$& $Z$& $Z$& $\cdot$& $\cdot$& $\cdot$& $\cdot$& $\cdot$& $\cdot$& $\cdot$& $\cdot$& $\cdot$& $\cdot$& $\cdot$& $\cdot$&\\
$\cdot$& $\cdot$& $\cdot$& $\cdot$& $\cdot$& $\cdot$& $\cdot$& $\cdot$& $Z$& $Z$& $Z$& $Z$& $\cdot$& $\cdot$& $\cdot$& $\cdot$& $\cdot$& $\cdot$& $\cdot$& $\cdot$&\\
$\cdot$& $\cdot$& $\cdot$& $\cdot$& $\cdot$& $\cdot$& $\cdot$& $\cdot$& $\cdot$& $\cdot$& $\cdot$& $\cdot$& $Z$& $Z$& $Z$& $Z$& $\cdot$& $\cdot$& $\cdot$& $\cdot$&\\
$\cdot$& $\cdot$& $\cdot$& $\cdot$& $\cdot$& $\cdot$& $\cdot$& $\cdot$& $\cdot$& $\cdot$& $\cdot$& $\cdot$& $\cdot$& $\cdot$& $\cdot$& $\cdot$& $Z$& $Z$& $Z$& $Z$&\\
$Z$& $\cdot$& $\cdot$& $Z$& $\cdot$& $\cdot$& $Z$& $Z$& $\cdot$& $\cdot$& $Z$& $Z$& $Z$& $\cdot$& $\cdot$& $Z$& $\cdot$& $\cdot$& $\cdot$& $\cdot$&\\
$\cdot$& $\cdot$& $\cdot$& $\cdot$& $Z$& $\cdot$& $\cdot$& $Z$& $Z$& $Z$& $\cdot$& $\cdot$& $\cdot$& $\cdot$& $Z$& $Z$& $Z$& $\cdot$& $\cdot$& $Z$&\\
$\cdot$& $Z$& $Z$& $\cdot$& $\cdot$& $\cdot$& $\cdot$& $\cdot$& $Z$& $\cdot$& $\cdot$& $Z$& $Z$& $Z$& $\cdot$& $\cdot$& $\cdot$& $\cdot$& $Z$& $Z$&\\
$\cdot$& $\cdot$& $Z$& $Z$& $\cdot$& $Z$& $Z$& $\cdot$& $\cdot$& $\cdot$& $\cdot$& $\cdot$& $\cdot$& $Z$& $Z$& $\cdot$& $Z$& $Z$& $\cdot$& $\cdot$&\\
$Z$& $Z$& $\cdot$& $\cdot$& $Z$& $Z$& $\cdot$& $\cdot$& $\cdot$& $Z$& $Z$& $\cdot$& $\cdot$& $\cdot$& $\cdot$& $\cdot$& $\cdot$& $Z$& $Z$& $\cdot$&\\
\hline
$\cdot$& $\cdot$& $X$& $X$& $\cdot$& $\cdot$& $\cdot$& $\cdot$& $X$& $\cdot$& $\cdot$& $X$& $X$& $\cdot$& $\cdot$& $X$& $\cdot$& $\cdot$& $\cdot$& $\cdot$& $=X_0$\\
$X$& $\cdot$& $\cdot$& $X$& $\cdot$& $\cdot$& $X$& $X$& $\cdot$& $\cdot$& $\cdot$& $\cdot$& $\cdot$& $\cdot$& $\cdot$& $\cdot$& $\cdot$& $\cdot$& $X$& $X$& $=X_1$\\
$Z$& $\cdot$& $\cdot$& $Z$& $\cdot$& $\cdot$& $Z$& $Z$& $\cdot$& $\cdot$& $\cdot$& $\cdot$& $\cdot$& $\cdot$& $\cdot$& $\cdot$& $\cdot$& $\cdot$& $Z$& $Z$& $=Z_0$\\
$\cdot$& $\cdot$& $Z$& $Z$& $\cdot$& $\cdot$& $\cdot$& $\cdot$& $Z$& $\cdot$& $\cdot$& $Z$& $Z$& $\cdot$& $\cdot$& $Z$& $\cdot$& $\cdot$& $\cdot$& $\cdot$& $=Z_1$\\
\end{tabular}
\end{minipage}

\caption{\textbf{Stabilizers and logical operators for the [[10,2,3]] and [[20,2,6]] codes.}}
\label{tab:codes}
\end{table*}

\begin{figure*}
    \centering
    \includegraphics[width=0.83\linewidth]{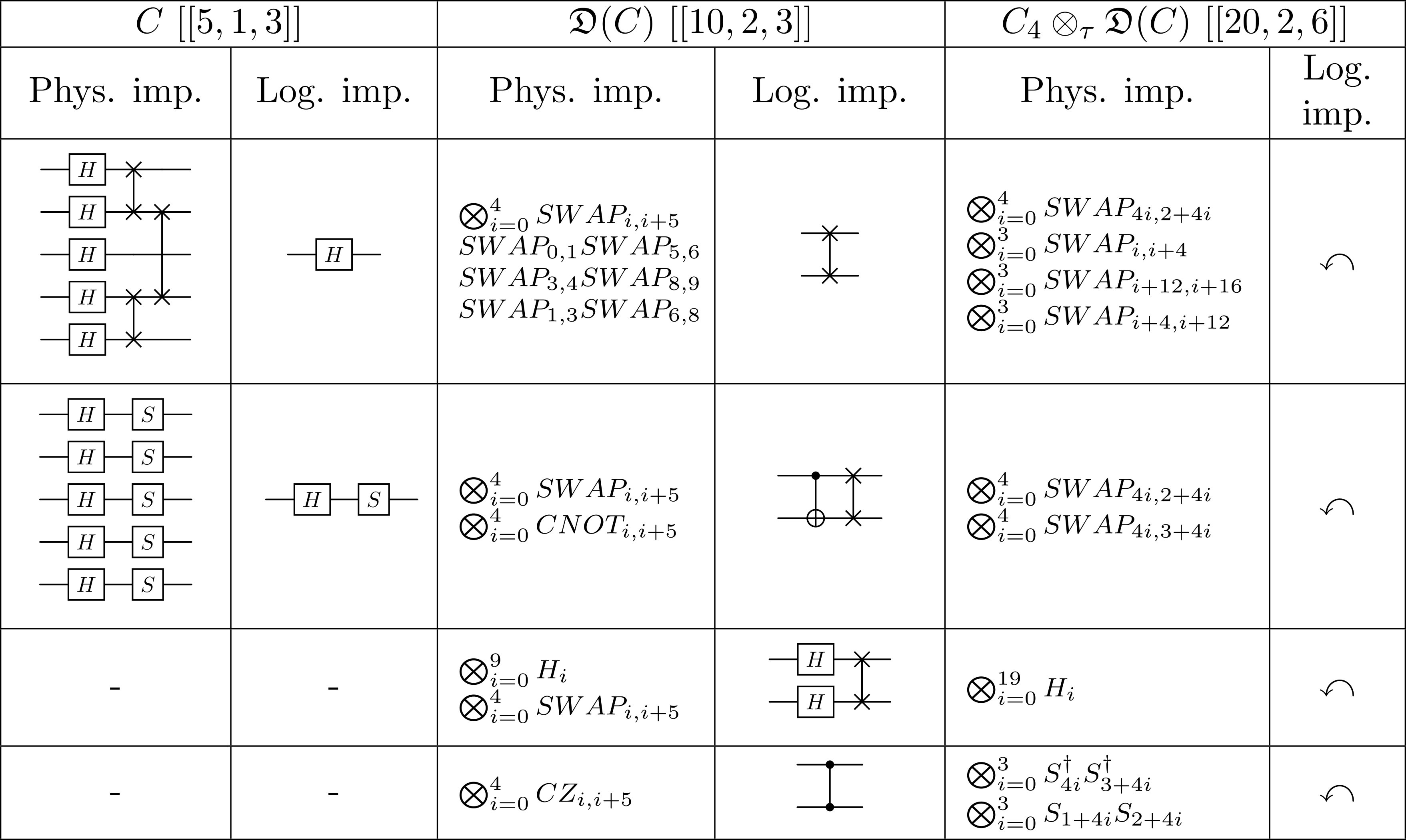}
    \caption{\textbf{Logical gates for the $C_4$-Helix code sequence}. SWAP-transversal gates on the non-CSS $[[5,1,3]]$ code lifting to SWAP- and fold-transversal logical gates on the $[[10,2,3]]$ symplectic double code, lifting again to SWAP-transversal gates on the $[[20,2,6]]$ CSD code.}
    \label{fig:actions}
\end{figure*}

\subsection{Logical Clifford gates}

The symplectic double and CSD constructions yield codes with many fold-transversal and SWAP-transversal (automorphism) gates, respectively, see Section 3.1 of Ref.~\cite{berthusen2025simple} for complete details. We list the available gates for the sequence of codes $[[5,1,3]] \rightarrow [[10,2,3]] \rightarrow [[20,2,6]]$ in Fig.~\ref{fig:actions}. Two gates come from lifting the $\overline{H}$ and $\overline{H} \overline{S}$ logical gates available on the $[[5,1,3]]$ code, while the H-SWAP and CZ-S gates~\cite{Breuckmann_2024} arise from the ZX-duality on the $[[10,2,3]]$ code. All of these gates are automorphisms on the $[[20,2,6]]$ code, and as such can be implemented using only single-qubit gates and qubit relabeling in software. These automorphism gates, $G$, generate a group of size $|G| = 36$. 

\begin{figure}
    \centering
    \includegraphics[width=\linewidth]{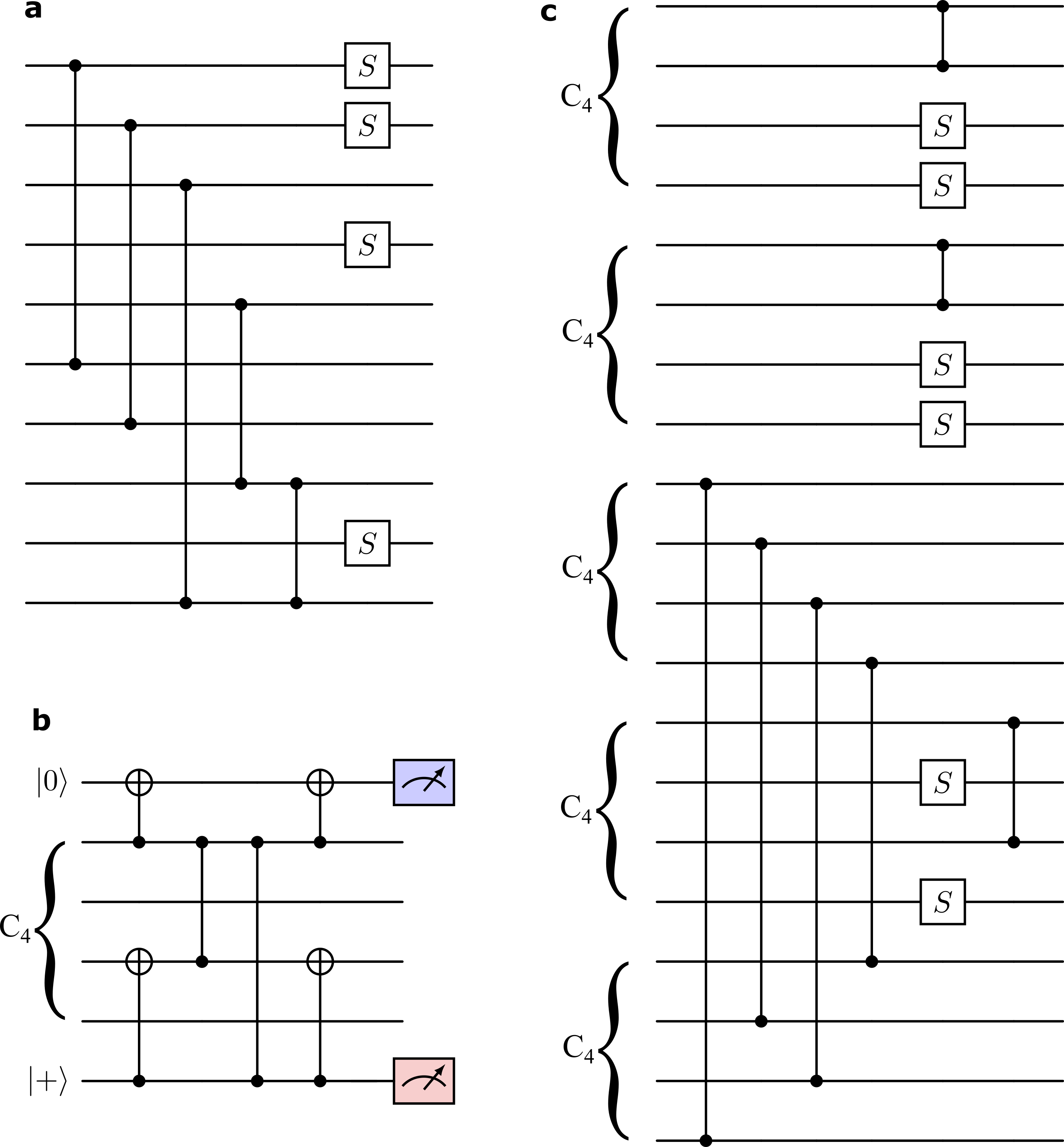}
    \caption{\textbf{Logical phase gates. a.} Implementation of $\overline{S}_0 \overline{S}_1$ on the $[[10,2,3]]$ symplectic double code. \textbf{c.} Non-fault-tolerant implementation of $\overline{S}_0 \overline{S}_1$ on the $[[20,2,6]]$ CSD code. All physical $CZ$ operations are not immediately fault-tolerant, even if followed by QEC. These operations are made fault-tolerant using the flag gadget shown in panel \textbf{b} and following with a round of QEC. }
    \label{fig:ft_phase_gate}
\end{figure}

To complete the two-qubit symplectic group, where $|\rm{Sp}_4(\mathbb{F}_2)| = 720$, an additional logical gate is required. We choose to implement a global $S$ gate, $\overline{S}_0 \overline{S}_1$. In Ref.~\cite{berthusen2025simple} it was suggested to realize this gate through gate injection~\cite{Zhou2000} during quantum error correction; however, the overheads of doing so are prohibitive, especially on near-term hardware. Instead, we consider an optimization-based approach~\cite{kuehnke2025} to search for gates implementing $\overline{S}_0\overline{S}_1$ on the $[[10,2,3]]$ code. We choose a gate that lifts nicely when implemented on the logical qubits of the five $[[4,2,2]]$ codeblocks; it and its lifted version are shown in Fig.~\ref{fig:ft_phase_gate}(a), (c). Unfortunately, due to the intra-block physical CZ gates, the circuit in Fig.~\ref{fig:ft_phase_gate}(c) is not immediately fault-tolerant. To remedy this, we replace each physical CZ gate with the flag gadget~\cite{Chao_2018} depicted in Fig.~\ref{fig:ft_phase_gate}(b). We verify with Stim~\cite{Gidney_2021} that, when followed by syndrome extraction, the circuit-level distance of this $\overline{S}_0\overline{S}_1$ gate is 6. Note that, as described, the logical global phase gate actually implements $\overline{S}_0 \overline{S}_1 \overline{Z}_0 \overline{Z}_1$ and the logical CZ gate actually implements $\overline{CZ}_{0,1} \overline{Z}_1$. While these actions are correct in the symplectic sense, it is required to determine if there are residual Paulis when implementing some two-qubit logical Clifford operation.

\begin{figure}
    \centering
    \includegraphics[width=\linewidth]{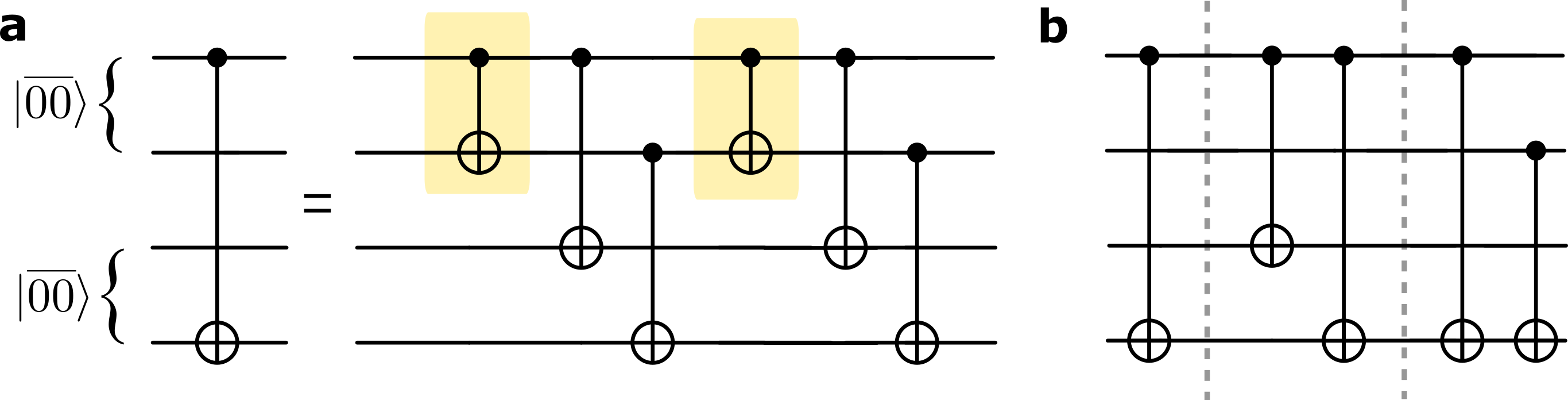}
    \caption{\textbf{Targeted interblock CNOT gate. a.} Circuit to implement a targeted inter-block CNOT gate using only transversal CNOT and automorphism gates (highlighted in yellow). \textbf{b.} Available inter-block gates via chain maps~\cite{benhemou2026}. Each gate/gate sequence (separated by a dashed line) can be implemented in depth-2 with a circuit-level distance of 6. Conjugating by SWAP gates in both/either codeblock further increases the available operations.}
    \label{fig:cnot-circuit}
\end{figure}

We can express any element of the two qubit symplectic group (and hence the full Clifford group by including transversal logical Pauli gates) with a sequence of gates from $G$ interleaved with the $\overline{S}_0 \overline{S}_1$ gate. Finding a valid sequence of logical generators then reduces to a word factorization problem in $\rm{Sp}_{2k}(\mathbb{F}_2)$. For arbitrary $k$, this is intractable to solve efficiently; fortunately, we can brute force search over $\rm{Sp}_4(\mathbb{F}_2)$ for factorizations that are optimal in the number of applied $\overline{S}_0 \overline{S}_1$ gates. We do this by performing Dijkstra's algorithm on a Cayley graph of $\rm{Sp}_4(\mathbb{F}_2)$ in which edges representing the global phase gate are assigned a nonzero cost and all other edges---corresponding to the free automorphism gates---have zero cost. Hence the optimal path through the graph is the one utilizing the fewest global phase gates. We then compute the shortest path between the identity and each element of $\rm{Sp}_4(\mathbb{F}_2)$ to build a lookup-table logical compiler. One such compiled Clifford is shown in Fig.~\ref{fig:clifford_bench_main}(d). The maximum number of global phase gates needed to implement an arbitrary two-qubit logical Clifford is three, and the full distribution is 0: 36, 1: 324, 2: 324, 3: 36.

Adding transversal inter-block CNOT gates to $G$ and $\overline{S}_0\overline{S}_1$ is sufficient to generate the full Clifford group on an arbitrary number of codeblocks. One important circuit identity that allows us to implement targeted inter-block CNOT gates is shown in Fig.~\ref{fig:cnot-circuit}(a). Alternatively, we can realize the CNOT circuits in Fig.~\ref{fig:cnot-circuit}(b) using fault-tolerant, depth-2 chain maps found using an automated chain map discovery tool~\cite{benhemou2026}. While the depths are the same between two methods, two rounds of error correction are required for the transversal approach---one after each transversal CNOT gate---whereas, only a single round of error correction is required following the chain map. Hence the cost of realizing inter-code operations via chain maps is effectively half that of the transversal approach; even larger savings are available when implementing logical FANIN/FANOUT gates. An arbitrary $m$-block Clifford operator is compiled into the $C_4$-Helix gateset by first expressing it in terms of inter-block CNOT/CZ gates (targeted and transversal) and intra-block Clifford operators, $U_i \in \mathcal{C}_2$ on each $C_4$-Helix codeblock. These $U_i$'s can then be expressed in terms of the logical generators. This process is illustrated in Fig.~\ref{fig:clifford_bench_main}(a)-(d). It may be possible to first recompile the circuit in such a way that expensive inter-block CNOT gates are minimized. Compiling an arbitrary $m$-block Clifford+T circuit can be done similarly, with the addition that non-Clifford gates also flush the accumulation of any intra-block Clifford operators.

\begin{figure}
    \centering
    \includegraphics[width=0.95\linewidth]{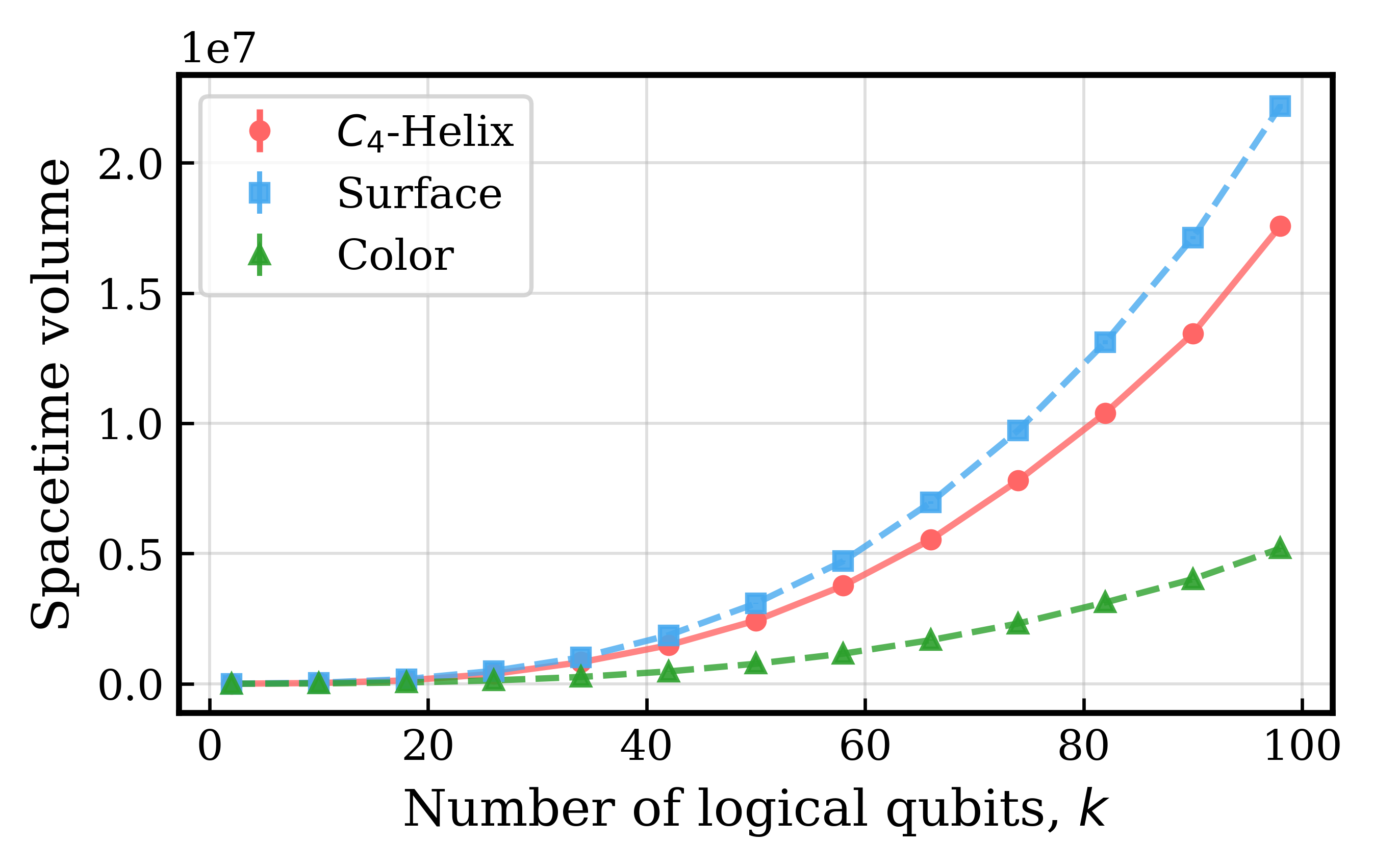}
    \caption{\textbf{Spacetime volume versus problem size}. Spacetime volume (mean compiled logical depth times number of physical qubits required to encode $k$ logical qubits) of uniformly random $k$-qubit Clifford operators, comparing the $C_4$-Helix gateset, the surface code, and the color code. Each operator is sampled uniformly at random and synthesized into a minimal physical-gate circuit, whose layers are then costed by the native operations of each architecture. Note that syndrome extraction overhead is not considered in this analysis. }
    \label{fig:cliff_bench}
\end{figure}

To illustrate the efficient logical compilation afforded by the $C_4$-Helix gateset, we compile uniformly random $k$-qubit Clifford operators and compare the resulting logical circuit depth against surface code and color code implementations of the same operators. For each qubit count, we sample random Clifford operators, synthesize a minimal physical-gate representation, and cost each layer according to the native operations of the respective architecture: inter-block CNOTs and intra-block $\mathcal{C}_2$ operators (via the lookup-table compiler) for the $C_4$-Helix code, fully transversal, depth-1 gates for the color code, and transversal gates---save for the depth-4 $S$ gate~\cite{chen2026}---for the rotated surface code. As shown in Fig.~\ref{fig:cliff_bench}, the $C_4$-Helix code, despite encoding $k>1$ logical qubits, achieves a spacetime volume (compiled depth times number of physical qubits) that beats the $d=5$ surface code and maintains a small constant factor between the color code. We again emphasize that uniformly random Clifford operators represent a worst-case workload for the $C_4$-Helix code. In practice, quantum algorithms can be compiled to rely predominantly on its inexpensive transversal and automorphism gates, avoiding the costly single-block Clifford and targeted inter-block CNOT operations that dominate the random-circuit cost. As such, the overhead reported here is an upper bound; for structured algorithms the practical cost would fall well below these values, substantially narrowing the gap with the surface and color codes.~\cite{LeBlond2026InPrep}.

\subsection{Syndrome extraction}
\label{apx:syndrome_extraction}

\begin{figure}[t]
    \centering
    \includegraphics[width=\linewidth]{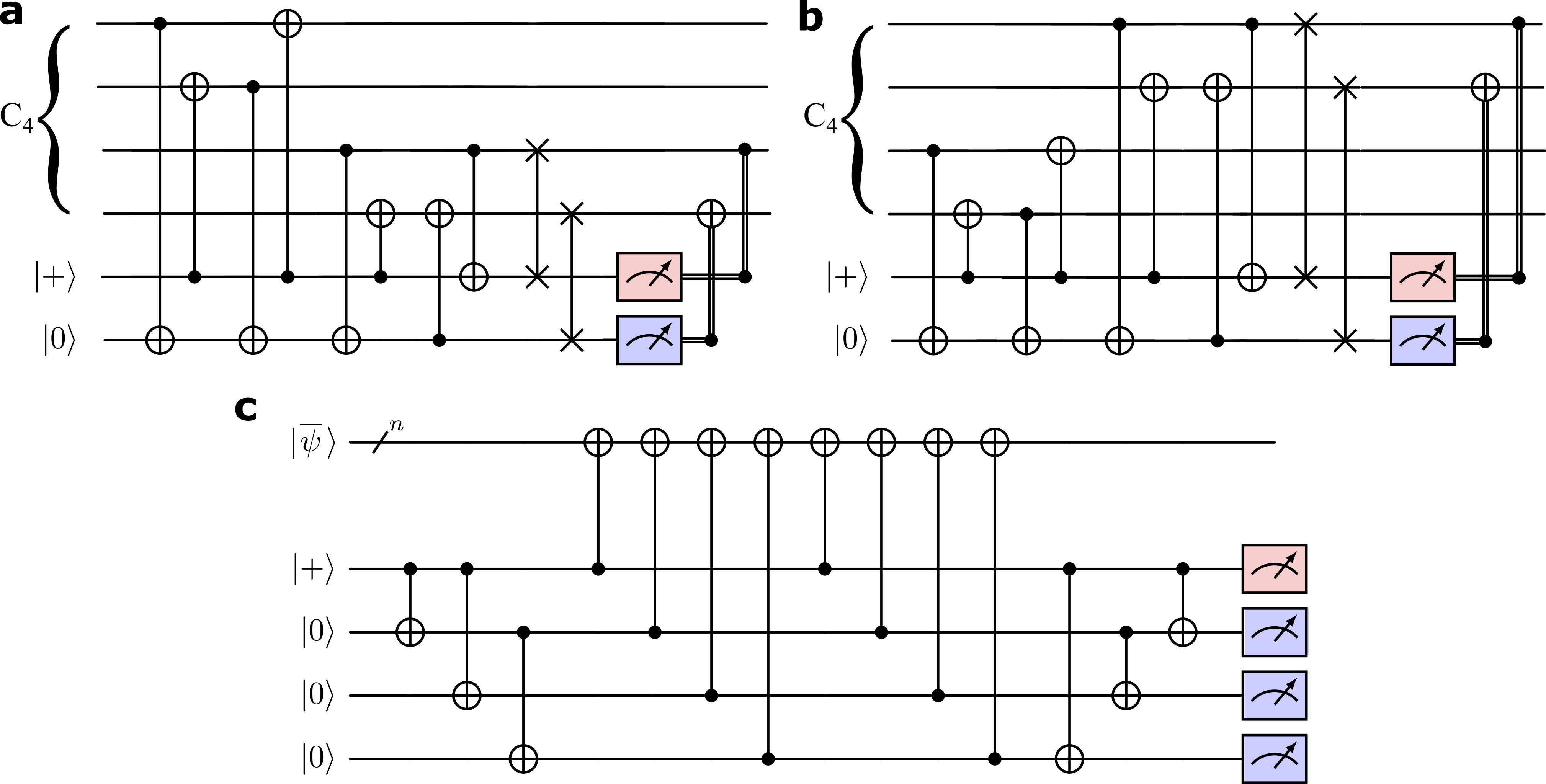}
    \caption{\textbf{Syndrome extraction circuits for the $C_4$-Helix code.} \textbf{a.} Circuit to measure the two stabilizer generators of the $C_4$ code. In order to mitigate the effects of leakage, each syndrome extraction round swaps out two data qubits for two fresh ancilla qubits; hence all four data qubits are exchanged every two rounds by alternating the circuits in panels \textbf{a} and \textbf{b}. The SWAP gates drawn are implemented virtually through qubit relabeling and do not introduce additional error. \textbf{c.} Modified Shor-style syndrome extraction for a single weight-8 stabilizer generator. }
    \label{fig:se_circuit}
\end{figure}

We use different syndrome extraction circuits to measure the stabilizer generators from different levels of concatenation: in a full syndrome extraction round, we first measure the lower-level, weight-$4$ stabilizers in parallel using the circuits shown in Fig.~\ref{fig:se_circuit}(a)-(b). We then measure the upper-level, weight-$8$ $X$-type stabilizers followed by the upper-level, weight-8 $Z$-type stabilizers using the circuit shown in Fig.~\ref{fig:se_circuit}(c). We use $32$ ancillas to parallelize the syndrome extraction as much as possible, resulting in a depth-$10$ circuit for measuring a complete set of $18$ stabilizers. 

To measure the lower-level stabilizer generators of the $C_4$ codeblocks, we use a circuit similar to Ref.~\cite{reichardt2025}, as shown in Fig.~\ref{fig:se_circuit}(a). It uses two ancilla qubits to measure the $X$- and $Z$-type stabilizer of a $C_4$ block in parallel. Each ancilla also acts as a flag qubit for the other ancilla to catch potential hook errors. Lastly, it swaps out two data qubits for the two fresh ancilla qubits, mitigating the effect of leakage on data qubits~\cite{aliferis2007fault, suchara2015leakage}. To ensure all data qubits get swapped out, we alternate between two slightly different circuits (see Fig.~\ref{fig:se_circuit}(b)), where the difference is that they swap out different qubits of the $C_4$ code. By doing so, all data qubits get swapped out after two rounds of syndrome extraction. Since different lower-level $C_4$ blocks are supported on different physical qubits, all lower-level stabilizer can easily be measured in parallel. To measure the upper-level, weight-$8$ $X$-type stabilizers, we use the circuit shown in Fig.~\ref{fig:se_circuit}(c), which uses four ancilla qubits (one syndrome qubit plus three flag qubits) per stabilizer and has CNOT depth of $6$. It is possible to find a schedule to measure all four weight-$8$ $X$-type checks in parallel. Since the code is self-dual, the circuit for measuring the corresponding $Z$-type checks can be obtained by simply reversing the direction of CNOTs and modifying the initialization and measurement bases for ancilla qubits accordingly. 

Since we have enough ancilla qubits at our disposal, we use different ancillas for measuring the upper-level $X$- and $Z$-type stabilizers; this allows us to start the $Z$ syndrome measurement circuits before the $X$ syndrome measurement circuits have finished and hence lower the total depth. Moreover, the upper-level $X$ syndrome extraction circuit can start before the lower-level syndrome extraction circuit has finished to further lower the total depth to 10. 

\subsubsection{Adaptive syndrome extraction}
\label{section:ase}

Since the $C_4$-Helix code is a concatenated code with the $C_4$ code as the lower code, we can readily apply adaptive syndrome extraction (ASE)~\cite{Berthusen_2025} to reduce spacetime overheads and consequently lower logical error rates. In the original prescription of ASE, the stabilizer generators of the lower $C_4$ codeblocks were first measured, then only concatenated generators that shared support with a triggered $C_4$ generator are measured. For the $C_4$-Helix code, each $C_4$ generator overlaps with four out of the five concatenated generators coming from the twisted toric code. Measuring the exact set of overlapping generators adds complexity to implementation and decoding, see below, that is likely not worth the slight reduction in gate count. As such, we make the simplification that if any of the $C_4$ stabilizer generators indicate that an error has occurred, all ten weight-8 checks (both $Z$ and $X$) are measured. 

We can also apply ASE in a setting where logical gates are applied, with some modifications to account for dangerous undetectable errors. In particular, the composition of the logical phase gate with automorphism gates can cause high-weight errors which are missed by the $C_4$ generators. Consider Fig.~\ref{fig:ft_phase_gate}(c) and suppose we have some weight-2 $XX$ error as an input to the third $C_4$ block. This $XX$ error propagates along the CZ gates to a $ZZ$ error on the fifth $C_4$ block. Both of these errors are undetectable by the underlying $C_4$ checks as they have even weight, and so no upper-level generators are scheduled to be measured. Following the application of some automorphism, e.g., the first two columns of Fig.~\ref{fig:actions} and another $\overline{S}_0\overline{S}_1$ gate, the same $XX$ error can be propagated to a $ZZ$ error on a different $C_4$ codeblock. We resolve this with the following decision tree: if no $C_4$ syndromes are observed, we measure the two upper-level, $X$-type stabilizers which have support on the two suspect $C_4$ codeblocks; if any $C_4$ stabilizer has a $-1$ measurement result, we measure all upper-level stabilizers. Doing so catches dangerous weight-2 errors while still reducing gate count/circuit depth on average.

To decode an experiment with $r$ rounds of adaptive syndrome extraction, we must compute the DEM of the exact circuit that ran out of the $2^r$ possible paths depending on whether or not the concatenated generators were measured each round (see Appendix~\ref{apx:simulations} for explicit instructions on constructing the DEMs for both memory and logic).

\subsection{Non-Clifford gates}
\label{apx:non-clifford}

As discussed in the main text, we can efficiently get magic $\ket{T}$ resource states into the $[[20,2,6]]$ $C_4$-Helix code by consuming $|T\rangle$ resource states from a surface code block using a homomorphic CNOT gate. However, a universal gate set is also achievable using only the $C_4$-Helix code. Namely, we can implement a magic $\ket{CZ}$ state: $\ket{CZ} = (\ket{00} + \ket{10} + \ket{01})/{\sqrt{3}}$, using zero-level distillation~\cite{Itogawa_2025}. Instead of preparing a noisy $\ket{CZ}$ state to distill, we can prepare the easily accessible $\ket{++}$ logical state. With probability 3/4, we successfully project the $\ket{++}$ state into the $\ket{CZ}$ state.

 The probability of obtaining a -1 measurement outcome is then 1/4, and the resulting state will be $\ket{11}$, a non-magic state. In this case, the zero-level distillation has failed and must be reattempted. To restart the process, one can apply $X_0X_1$ followed by a transversal Hadamard state to convert from the post-measurement $\ket{11}$ state to $\ket{++}$. One other consideration is the method in which the $CZ$ operator is measured: using a 20-qubit cat state is fault-tolerant, but requires significant overhead to prepare. Alternatives might be to use bare ancillas with appropriate flag gadgets~\cite{forlivesi2025}, or the modified Shor-style syndrome extraction described in Section \ref{apx:syndrome_extraction}.

\begin{figure}[t]
    \centering
    \includegraphics[width=\linewidth]{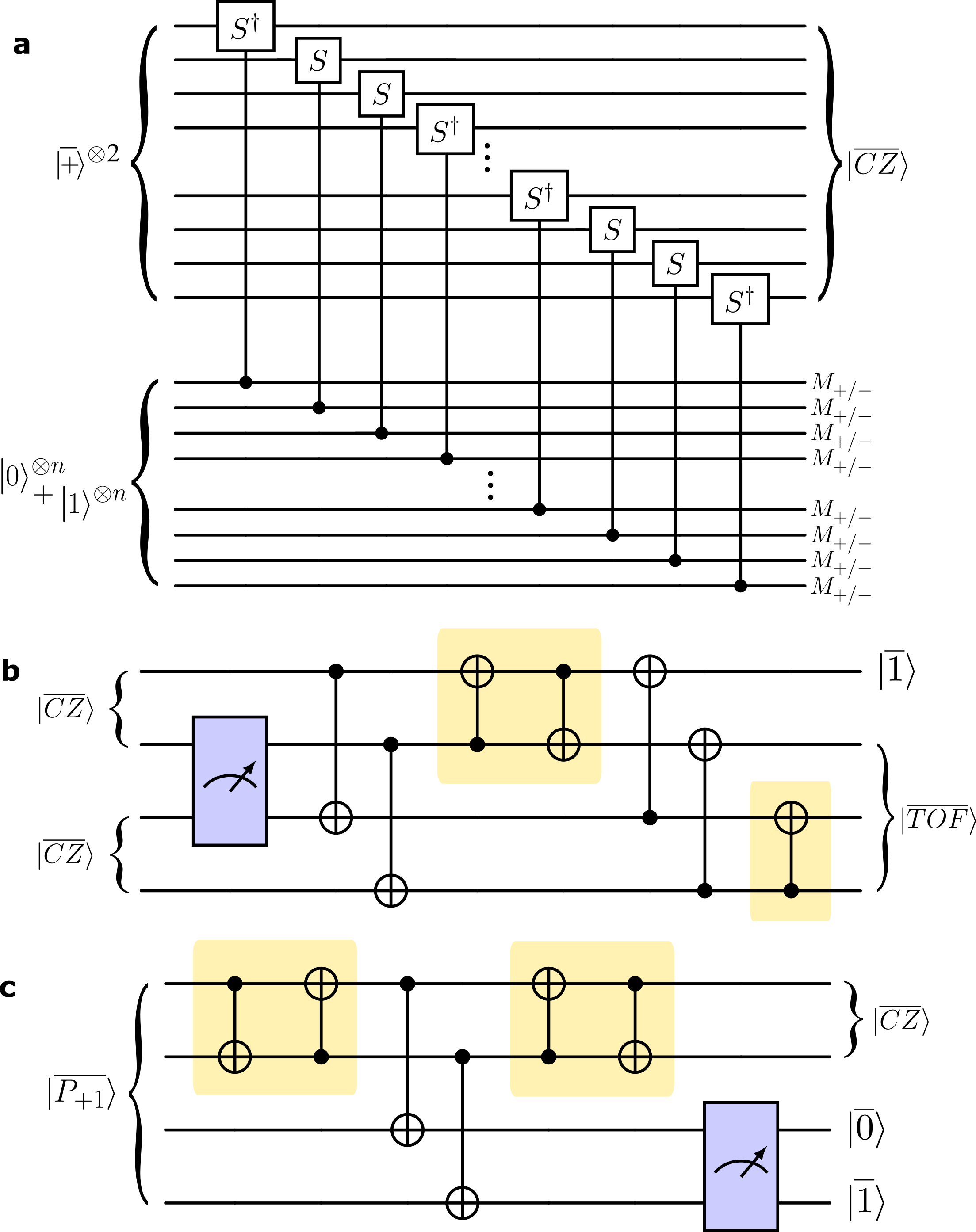}
    \caption{\textbf{Process for distilling a $\ket{\rm TOF}$ state on the $C_4$-Helix code. a.} Circuit to zero-level distill the $\ket{CZ}$ state using the input state $\ket{+}^{\otimes n}$. Here we have depicted the ancilla as being a large cat state, $\ket{0}^{\otimes n} + \ket{1}^{\otimes n}$; however, it may be feasible to use a flagged circuit to fault-tolerantly extract the measurement result. \textbf{b.} Circuit to take two $\ket{CZ}$ states to $\ket{1}\ket{\rm TOF}$ using a $Z_2Z_3$ measurement, postselecting on the -1 outcome, and applying transversal interblock CNOT gates and SWAP-transversal intrablock CNOT gates. The circuit is immediately fault-tolerant (with inserted error correction) except for the $Z_2Z_3$ measurement. This can be made fault-tolerant with the appropriate ancillas and flags. \textbf{c.} Circuit to take the projected +1 measurement result to one that recovers one of the two input $\ket{CZ}$ states using transversal interblock CNOT gates and SWAP-transversal intrablock CNOT gates. The $Z_3Z_4$ measurement is conditioned on obtaining a -1 measurement outcome.}
    \label{fig:cz_to_tof}
\end{figure}

Following Ref.~\cite{Gupta2024}, we can use two $\ket{CZ}$ states to create a Toffoli state,
\begin{equation}
    \ket{\rm TOF} \propto  \ket{000} + \ket{010} + \ket{100} + \ket{111}
    \label{eq:toff}
\end{equation}

The procedure to do so is shown in Fig.~\ref{fig:cz_to_tof}. 
We first measure the operator $Z_2Z_3$ on $\ket{CZ}^{\otimes 2}$, which has the following action on the input state
\begin{equation}
    Z_2Z_3\ket{CZ}^{\otimes 2} \propto \sum_{\text{9 terms}}(-1)^{b+c}\ket{abcd}
    \label{z2z3}
\end{equation}
The +1 subspace is spanned by the five terms with eigenvalue +1
\begin{align}
    \ket{P_{+1}} &\propto \ket{0000} + \ket{0001} + \ket{0110} + \ket{1000} + \ket{1001},
\end{align}
while the -1 subspace is spanned by the remaining four terms with eigenvalue -1,
\begin{align}
    \ket{P_{-1}} &\propto \ket{0010} + \ket{0100} + \ket{0101} + \ket{1010}.
\end{align}

After the $Z_2Z_3$ measurement, the states $\ket{P_{+1}}$ and $\ket{P_{-1}}$ are obtained with probability 5/9 and 4/9, respectively. In the result of a -1 measurement outcome, the resulting state $\ket{P_{-1}}$ is in a convenient form to obtain the $\ket{\rm TOF}$ state, Eq.~\eqref{eq:toff}. By applying the remaining fault-tolerant circuit in Fig.~\ref{fig:cz_to_tof}(b) consisting of only transversal CNOT gates between $[[20,2,6]]$ codeblocks and automorphism gates within a codeblock, we obtain the state $\ket{1}\ket{\rm TOF}$. In the event of a +1 measurement result, we project into the $\ket{P_{+1}}$ state, which is not of a usable form; however, we can apply the circuit depicted in Fig.~\ref{fig:cz_to_tof}(c), again consisting of only transversal CNOT gates and automorphism gates, to attempt to recover one of the input $\ket{CZ}$ states. The state after the CNOT gates in Fig.~\ref{fig:cz_to_tof}(c) is 
\begin{equation}
    R_{\rm CNOT}\ket{P_{+1}} \propto \ket{0000} + \ket{1000} + \ket{0001} + \ket{1001} + \ket{0101}.
    \label{eq:pre_recovery}
\end{equation}
On this state we then measure the $Z_3Z_4$ operator. Similar to Eq.~\eqref{z2z3}, we have
\begin{equation}
    Z_3Z_4 R_{\rm CNOT}\ket{P_{+1}} \propto \sum_{\text{5 terms}} (-1)^{c+d} \ket{abcd}
\end{equation}
It can be seen that the -1 subspace is spanned by the three terms in which have the form $\ket{ab01}$. This measurement result occurs with probability 3/5, and the resulting state is $\ket{CZ}\ket{01}$. Hence we have recovered one of the two input $\ket{CZ}$ states with which we can attempt another Toffoli conversion.

Now let us estimate the resources required to successfully prepare a Toffoli state. In particular, let us estimate the number of $\ket{CZ}$ state preparation attempts (via Fig.~\ref{fig:cz_to_tof}(a)) that are needed to yield one successfully prepared Toffoli state. For simplicity, let us first assume that the input $\ket{CZ}$ states can be prepared deterministically. We then have the following costs:
\begin{itemize}
    \item Round 1: We always require 2 inputs.
    \item Retry rounds: A retry is needed with probability 5/9.
    \begin{itemize}
        \item Recover a $\ket{CZ}$ state: We recover a single $\ket{CZ}$ state from  Eq.~\eqref{eq:pre_recovery} with probability 3/5.
        \item Recover nothing: We have to create two new $\ket{CZ}$ states with probability 2/5.
    \end{itemize}
\end{itemize}

The number of times we expect to have to repeat the $Z_2Z_3$ measurement is independent of the success probability of preparing a $\ket{CZ}$ state. We project into the $\ket{P_{-1}}$ state with probability 4/9, so we expect to have to do this 1/(4/9) = 9/4 times. Hence the expected number of \textit{retries} is $E_r$ = 9/4 - 1 = 5/4. The expected cost per retry depends on the success probability of recovering a $\ket{CZ}$ state from $\ket{P_{+1}}$. Given a retry, there is a 3/5 probability we only need one additional new input $\ket{CZ}$ state, and there is a 2/5 probability we need two new input states. Hence each retry after the initial round has an expected cost of $E_c = 7/5$. The total expected inputs (assuming deterministic preparation) is then
\begin{align}
    E_s &= (\text{Cost of Round 1)} + E_r \times E_c \nonumber \\ 
        &= 2 + \frac{5}{4} \times \frac{7}{5} = \frac{15}{4}.
        \label{eq:expected_cost}
\end{align}
And so if the input $\ket{CZ}$ states could be prepared deterministically, we would expect to need 15/4 = 3.75 input $\ket{CZ}$ states on average to successfully prepare a single $\ket{\rm TOF}$ state. Let us now incorporate the fact that the probability of successfully preparing a $\ket{CZ}$ state is less than one (3/4 for the instance where the state being measured is $\ket{++}$). We can just take the expected cost of Eq.~\eqref{eq:expected_cost} and multiply by the cost per successful $\ket{CZ}$ preparation. If a $\ket{CZ}$ state is prepared with probability $p$, then the expected cost is $1/p$, and so the final cost is simply $E_C = 15/(4p)$. At $p=3/4$, this works out to needing, on average, five $\ket{CZ}$ preparation attempts per successful $\ket{\rm TOF}$ state creation. Note that this does not mean we need five $[[20,2,6]]$ codeblocks. We can reuse the codeblocks in instances where the distillation fails, so we still only need two codeblocks. Of course, the retries could be sped up if additional $\ket{CZ}$ states were already on hand. This result also indicates that there is not much benefit to switching from $\ket{++}$ to $\ket{CZ}$ for the zero-level distillation step; we only save, on average, 1.25 input $\ket{CZ}$ states at the cost of requiring an expensive arbitrary state encoder circuit.

\section{Details of the $[[60,2,12]]$ and $[[100,2,18]]$ codes}
\label{apx:big_$C_4$-Helix}

\begin{figure*}
    \centering
    \includegraphics[width=0.9\linewidth]{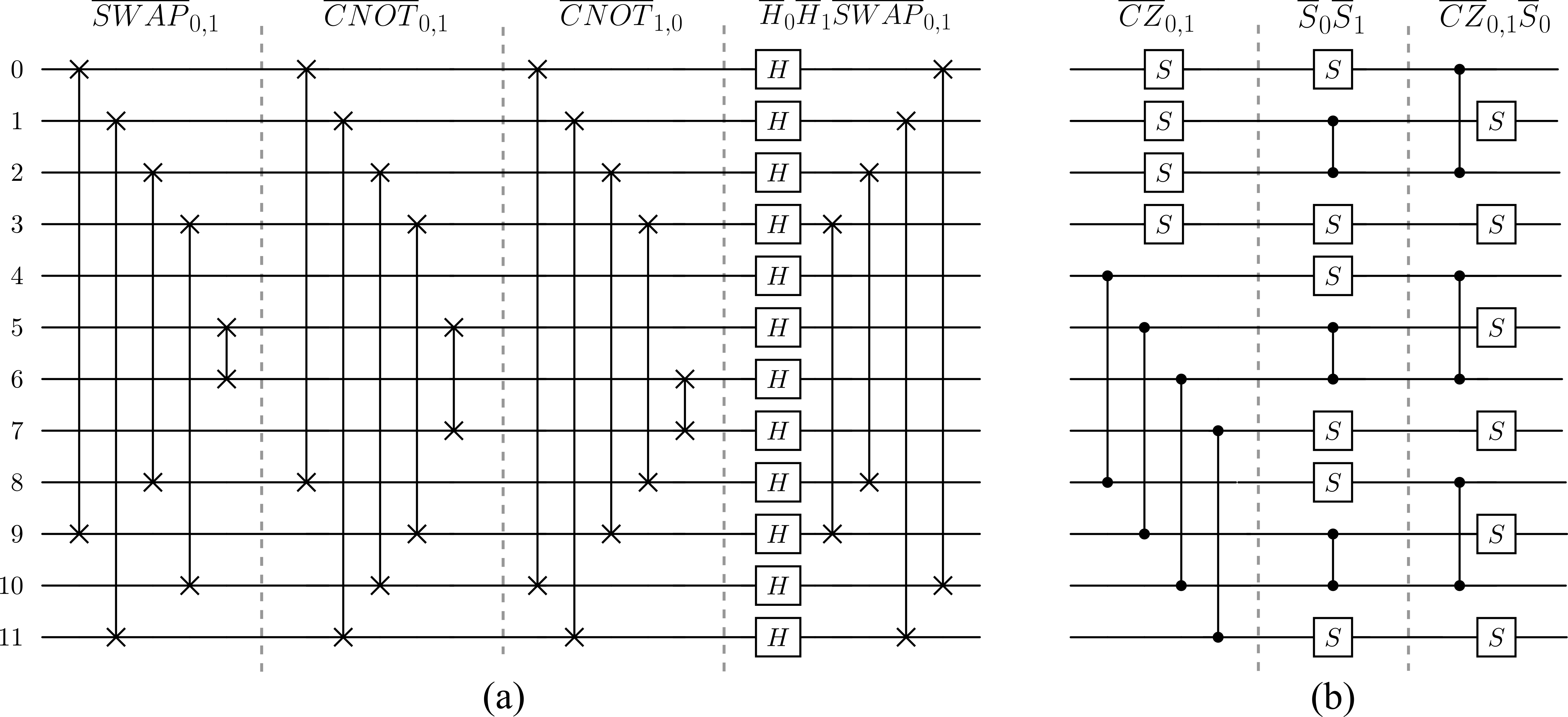}
    \caption{\textbf{Logical operators for the $[[12,2,4]]$ Carbon code. a.} SWAP-transversal gates implementable using only single qubit gates and qubit relabeling. \textbf{b.} Depth-1 fold transversal gates. Note that as compared to the $[[4,2,2]]$ code, the operator $\overline{CZ}_{0,1}$ is no longer SWAP-transversal.}
    \label{fig:carbon_gates}
\end{figure*}

In this Appendix, we provide the details for larger versions of the $[[20,2,6]]$ $C_4$-Helix code, namely the $[[60,2,12]]$ \textit{Carbon-Helix} code and $[[100,2,18]]$ \textit{Double-Helix} code. As with the $C_4$-Helix code,  Appendix~\ref{apx:code_construction}, we proceed through the CSD construction until we have obtained the $[[10,2,3]]$ twisted toric code. Then, instead of concatenating along the ZX-duality $\tau: \{i,i+5\}_{0,...,9}$ with the $[[4,2,2]]$ code, we use a code with a larger distance. Using the $[[12,2,4]]$ Carbon code~\cite{paetznick2024}, the stabilizer generators and logical operators of which are given in Table IV of Ref.~\cite{paetznick2024}, yields the $[[60,2,12]]$ Carbon-Helix code. Using the $[[20,2,6]]$ $C_4$-Helix code as the lower code yields the $[[100,2,18]]$ Double-Helix code.

These code parameters are already quite competitive: the $[[60,2,12]]$ code is $4\times$ more efficient than the $[[144,1,12]]$ rotated surface code and only $2.5\times$ worse than the $[[144,12,12]]$ gross code~\cite{bravyi2023highthreshold}. Similarly, the $[[100,2,18]]$ code is $6\times$ more efficient than the $[[324,1,18]]$ rotated surface code and uses only $2\times$ more qubits than the $[[288,12,18]]$ two-gross~\cite{yoder2025} code. The benefit with this construction over these higher-rate codes is that logical computation remain simple. 

Recall that implementing logical gates on the $C_4$-Helix family can be thought of performing the logical gates of the $[[10,2,3]]$ twisted toric code on the logical qubits of the lower-level ($[[4,2,2]], [[12,2,4]], [[20,2,6]]$) codes. Thus for computation to be convenient, the required `physical' gates listed in the third column of Table~\ref{fig:actions}, $\overline{SWAP}_{i,i+5}, \overline{CNOT}_{i,i+5}$, etc., as well as in Fig.~\ref{fig:ft_phase_gate}(a), $\overline{CZ}_{i, i+5} \overline{S}_i, \overline{S}_i \overline{S}_{i+5}$, etc., must be easy to implement on the lower-level code. The $[[4,2,2]]$ code satisfies this property. Fig.~\ref{fig:carbon_gates} lists the physical circuits that implement the required logical gates on the Carbon code. The final three CZ gates of Fig.~\ref{fig:ft_phase_gate}(a) can be realized between two Carbon codeblocks with a transversal CZ gate $\{(3,2),(2,3),(8,7),(7,8),(11,10),(10,11)\}$ (unlisted indices are $i \rightarrow i$) conjugated by logical SWAPs on both codeblocks. Compared to the $[[4,2,2]]$ code, only the $\overline{CZ}_{0,1}$ gate is no longer SWAP-transversal. And so, we now have two logical gates on the Carbon-$C_4$-Helix code that require physical two-qubit gates (and hence follow-up QEC), $\overline{CZ}_{0,1}$ and $\overline{S}_0 \overline{S}_1$. Just as we did for the $C_4$-Helix code, we can build a lookup-table logical compiler that minimizes the number of QEC cycles required. The maximum number of QEC cycles needed to implement an arbitrary two-qubit logical Clifford on the Carbon-$C_4$-Helix code is four, and the full distribution is 0: 12, 1: 96, 2: 252, 3: 324, 4: 36.

Computation for the $[[100,2,18]]$ Double-Helix code follows analogously by applying combinations of $[[20,2,6]]$ gates in fifth column of Table~\ref{fig:actions} and Fig.~\ref{fig:ft_phase_gate}(c). All gates in Table~\ref{fig:actions} remain SWAP-transversal, and so we can reuse the logical compiler from the $C_4$-Helix code that minimizes the number of $\overline{S}_0 \overline{S}_1$ applications.


\section{Simulations}
\label{apx:simulations}

Throughout the paper, we perform circuit-level simulations with Stim~\cite{Gidney_2021} as well as Guppy~\cite{koch2025} and Selene~\cite{selene_2025}. We primarily use Guppy and Selene to estimate error rates for the Quantinuum Helios trapped-ion quantum processor, and we use Stim to probe the low error-rate regime. We use the following circuit-level noise model:
\begin{itemize}
    \item Qubit reset operations have a probability $p$ of preparing the $\ket{1}$ state instead of the $\ket{0}$ state.
    \item Qubit measurement operations have probability $p$ of yielding a flipped result.
    \item The single-qubit depolarizing channel is applied after single-qubit gates with probability $p/10$.
    \item The two-qubit depolarizing channel is applied after two-qubit gates with probability $p$.
    \item The single-qubit dephasing (Z) channel is applied during idle operations with probability $p/10$.
\end{itemize}
Note that when simulating Helios, we instead adopt a device-specific stochastic Pauli error model derived from  benchmarking data~\cite{Ransford2026}.

\subsection{Defining Detectors and Sign Tracking}
We perform circuit-level decoding over the entire circuit with a detector error model (DEM) generated from Stim. In a standard memory experiment, rows of this matrix, referred to as detectors, are formed from parities of time-adjacent syndromes, $D_i^{(t)} = \sigma_i^{(t)} \oplus \sigma_i^{(t-1)}$, where $\sigma_i^{(t)}$ denotes the $i^{\rm th}$ syndrome measured at time $t$. In particular, since we prepare the logical state by projectively measuring the $Z$-type stabilizer generators (after initializing all physical qubits in the $\ket{+}$ state), the $Z$-syndromes will be initialized with random signs and it is necessary to make detectors out of parity difference rather than individual syndromes. Defining the detectors from the parity of time-adjacent syndromes (instead of the parity with respect to the random sign at initialization) would further help with localizing errors in space-time and thus can help with the decoding speed and accuracy. Here we are using space-time local loosely to mean local with respect to the graph representation of the circuit, since the checks for the $C_4$-Helix codes are geometrically non-local. In the presence of logical gates, such as in randomized logical Clifford benchmarking, building the detectors is more involved since one needs to track the random syndrome signs as they propagate from one stabilizer to another. While logical gates preserve the original codespace where all standard stabilizer generators have $+1$ sign, they do not necessarily preserve the image codespaces defined by random signs of the stabilizer generators resulting from projective state prep; Instead, the logical gates result in a deterministic linear mapping between the signs of the stabilizers before and after the logical gate. As such, $\sigma_i^{(t)}$ and $\sigma_i^{(t-1)}$ may no longer be related after applying some logical gate $\overline{U}$ at time $t$. We can determine how the current syndromes relate to the previous syndromes by using the transformation matrix $T$ that relates the stabilizer generators before and after the application of the logical gate.
\begin{equation}
    \mathcal{S} \mapsto \mathcal{S}'=T\mathcal{S}.
    \label{eq:stabs}
\end{equation}
Here, $\mathcal{S}$ is the  $\mathbb{F}_2^{(n-k) \times 2n}$ symplectic matrix describing the stabilizer generators of the code and $\mathcal{S}'$ is  their image in symplectic form under the logical gate $U$ and $T$ is the $\mathbb{F}_2^{(n-k) \times (n-k)}$ matrix relating them to each other. When applying multiple logical operations $\overline{U} = \overline{U}_i\overline{U_j}...\overline{U}_k$, between syndrome extraction rounds, either a total $T$ can be computed, or $T_i$'s for each $\overline{U}_i$ can be computed and composed $T = T_iT_j...T_k$. With $T$ computed, valid detectors are formed as
\begin{equation}
    D_i^{(t)} = \sigma^{(t)}_i \oplus (T^{-1}\sigma^{(t-1)})_i,
\end{equation}
since $T^{-1}\mathcal{S}$ gets mapped to $\mathcal{S}$ under $U$.
When using the image codespaces with non-trivial stabilizer signs, it is also necessary to track how the logical observables change under the application of logical gates. That is because in the presence of random stabilizer signs, logical operators which differ by a stabilizer are only equivalent up to a sign, making it necessary to track how the signs of the logical observables change as the result of the presence of random stabilizer signs as they propagate through logical gates. Just as we did for the stabilizer generators, we can keep track of the relative signs of $X$ and $Z$ logical operators $\overline{\sigma}_{X,j}^{(t)}$ and $\overline{\sigma}_{Z,j}^{(t)}$ for $j \in \{1,\cdots,k\}$ as they propagate through logical gates. We note that $\overline{\sigma}_{P,k}^{(t)}$ only tracks the relative signs that logical observables acquire under a logical Clifford gate and has nothing to do with the actual value of the logical operator. To track how $\overline{\sigma}_{P,j}^{(t)}\in \mathbb{F}_2^{2k}$ changes under a logical operation, we compute the action of the physical gate sequence on the symplectic representation of the $X$ and $Z$ logical operators: 
\begin{equation}\label{eq:ULTS}
    \mathcal{L}\mapsto \mathcal{L}^* = U \mathcal{L} \oplus T' \mathcal{S}.
\end{equation}
Here $T' \in \mathbb{F}_2^{2k \times (n-k)}$ represents the contribution from the stabilizer generators and $U\in \mathbb{F}_2^{2k \times 2k}$ represents the logical action of the Clifford gate $U$ in the symplectic form. Accordingly, the relative signs of the logical operators can be tracked as:
\begin{equation}
    \overline{\sigma}_{P,k}^{(t)} = (U^{-1}\overline{\sigma}^{(t-1)})_{P,k} \oplus (U^{-1} T' T^{-1}\sigma^{(t-1)})_{P,k}.
\end{equation}
This is because it follows from Eqs.~\eqref{eq:ULTS} and \eqref{eq:stabs} that $U^{-1}\mathcal{L}\oplus U^{-1} T' T^{-1}\mathcal{S}$ gets mapped to $\mathcal{L}$ under the corresponding Clifford operation. When measuring an observable at the end of the circuit, one should multiply the outcome by the relevant relative sign $\overline{\sigma}^{(t)}_{P,k}$ to cancel the sign that has been acquired due to random stabilizer signs. We set $\overline{\sigma}^{(0)}_{P,k}=+1$ at the start of the circuit.

We remark that the sign tracking can be entirely dropped if all stabilizers are initialized in their +1 eigenspace, e.g. by a unitary preparation circuit~\cite{forlivesi2025} or by applying destabilizers to any generators with an initial measured value of -1. In that case each syndrome can be a detector by itself. Those detectors however are not local in space-time, so decoders can still benefit from sign tracking of the syndromes as it results in space-time local detectors.

\subsection{Frontier decoding and confidence-based postselection}
\label{apx:decoder}

We decode the hardware experiments using the Frontier decoder~\cite{leverrier2026}, a pruned dynamic-programming decoder that approximates logical maximum-likelihood decoding. For a detector error model, let $x$ denote a fault configuration, $H$ the detector matrix, and $L$ the corresponding logical-observable matrix, such that
\begin{equation}
    Hx=s, \qquad Lx=\ell,
\end{equation}
where $s$ is the observed detector syndrome and $\ell$ labels the resulting logical class. The probability mass associated with a logical class is
\begin{equation}
    Z_\ell(s)
    =
    \sum_{\substack{x:Hx=s\\Lx=\ell}}
    \Pr(x).
\end{equation}
Logical maximum-likelihood decoding chooses the class with the largest $Z_\ell(s)$.

Frontier approximates these logical-class masses by processing fault variables sequentially while retaining a small set of boundary states. Each boundary state records the residual syndrome on the active detector boundary together with its accumulated logical label. States with identical boundary data are merged by summing their probability masses, after which low-scoring states are pruned. At the end of the circuit, the retained states are aggregated according to logical class, giving approximate masses $\widehat{Z}_\ell(s)$, and the decoder returns
\begin{equation}
    \ell^*
    =
    \text{arg max}_\ell \widehat{Z}_\ell(s).
\end{equation}
For the hardware experiments and for the simulations below, we apply Frontier directly to the detector error model of the entire circuit volume. However, we note that, through its narrow-frontier construction, Frontier naturally supports streaming decoding, for which memory requirements and latencies need not grow with the total circuit depth. We leave it to future work to develop the streaming version and to interface it with a quantum computer.

Since even distance codes can detect higher weight errors than can be reliably corrected, it is (optionally) possible to improve their performance through post-selection on detection of high weight errors.
Forced gap decoding~\cite{gidney2023, lee2026, xie2026, wills2026} is a general post-selection technique which can potentially improve the success probability of the decoding task at the expense of a tunable post-selection rate. Given a decoder $D$, syndrome $s$, and a set of observables $(o_1,\cdots,o_k)$, let $E^*=D(s)$ be the error that the decoder $D$ finds when decoding the syndrome $s$ and let $l^*\in \mathbb{Z}_2^{k}$ be the vector that specifies which observable value is flipped by $E^*$. Let $C(E^*)$ denote the cost of $E^*$, which is a measure of how unlikely it is for this error to happen. One can quantify the decoder's confidence in its solution by looking a the difference between $C(E^*)$ and $C(\widetilde{E}^*)$ where $\widetilde{E}^*$ is the error that the decoder would have found if it was restricted to solutions for which $\widetilde{l}^*\ne l^*$. For example if $E^*$ is the most likely error, $\widetilde{E}^*$ would be the most-likely error which is not logically equivalent to $E^*$. If $C(\widetilde{E}^*) \approx C(E^*)$, the decoder is not very confident on the solution it has found, while $C(\widetilde{E}^*) \gg C(E^*)$ means that $E^*$ is a high confidence solution. Forced-gap postselection rejects shots for which this separation falls below a chosen threshold.

For decoders that return only a single preferred correction, determining $\widetilde{E}^*$ generally requires additional constrained decoding calls in which the logical outcome is forced to differ from $\ell^*$. Frontier makes this comparison directly available: because its terminal output retains probability mass resolved by logical class.
We define the Frontier confidence gap as
\begin{align}
    \Gamma(s)
    &=
    \log \widehat{Z}_{\ell^*}(s)
    -
    \max_{\ell\neq\ell^*}
    \log \widehat{Z}_{\ell}(s) \\
    &=
    \log
    \frac{\widehat{Z}_{\ell^*}(s)}
    {\widehat{Z}_{\ell^{(2)}}(s)},
\end{align}
where $\ell^{(2)}$ denotes the second-most-likely logical class. Larger values of $\Gamma$ indicate greater separation between the decoder's preferred logical class and its nearest competitor. For forced-gap postselection, we choose a threshold $\gamma$ and reject a shot whenever $\Gamma(s)<\gamma$. The threshold $\gamma$ can be tuned to trade logical error rate against post-selection overhead.

\subsection{Low error-rate simulations}
\begin{figure}[t]
    \centering
    \includegraphics[width=0.95\linewidth]{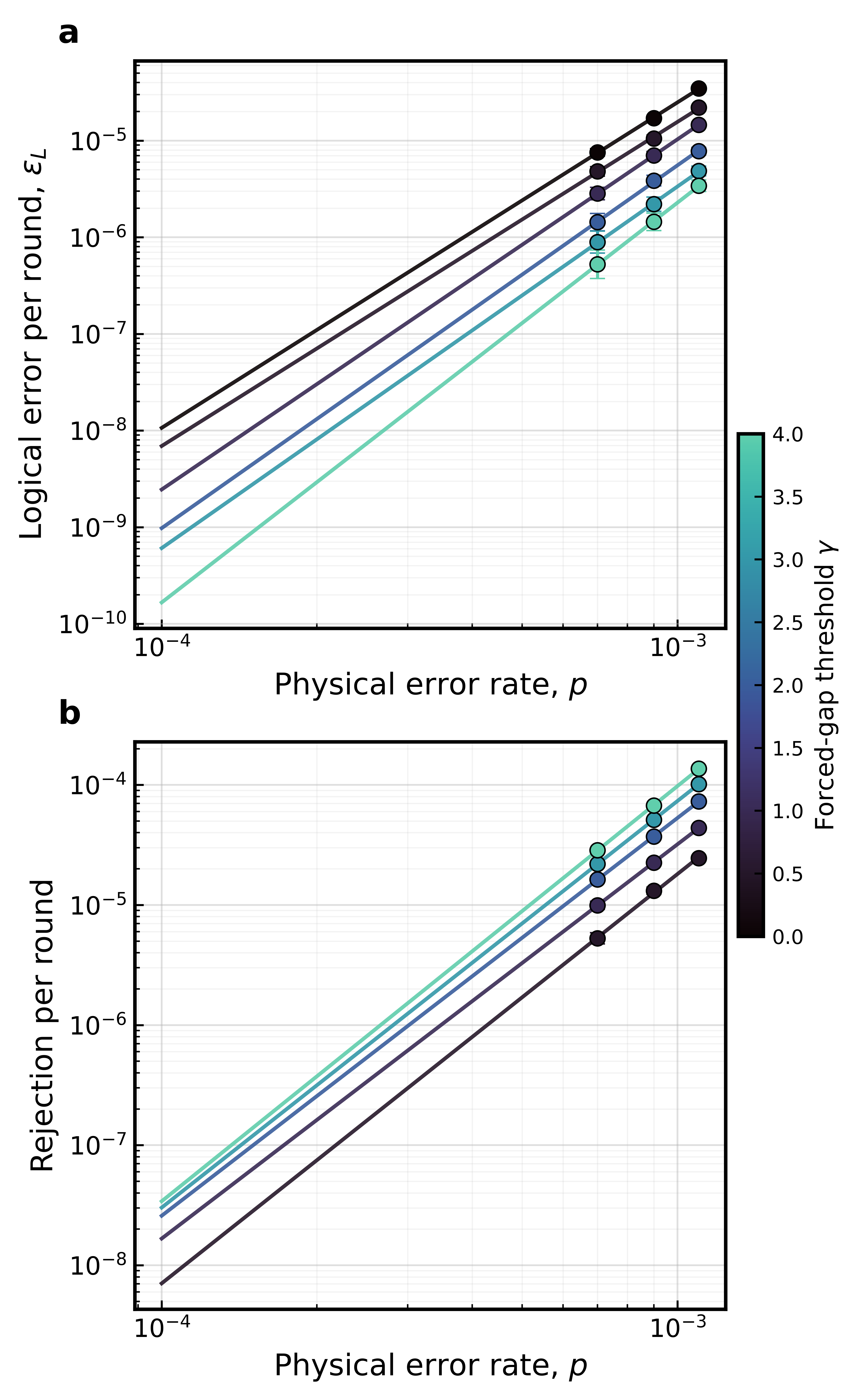}
    \caption{\textbf{Low physical error rate memory experiment simulations.} Memory simulations of the $[[20,2,6]]$ code as prepared using projective measurement and decoded using Frontier over the entire $r=12$ rounds of syndrome extraction. \textbf{a.} Logical error rate per round as a function of the physical error rate. \textbf{b.} Postselection rate per round as a function of the physical error rate. For $\gamma=0$, no postselection is performed. Postselection intensity ($\gamma$) is indicated by the line color.}
    \label{fig:memory_frontier}
\end{figure}
In this section, we present additional circuit-level simulations extended to physical error rates as low as $p= 10^{-4}$--- a regime motivated by two-qubit gate infidelities already demonstrated in trapped-ion testbeds~\cite{hughes2025}. Despite the \chelix{} code only having a distance of six, we find that, when combined with modest postselection from the Frontier decoder, logical error rates per round near $10^{-10}$ are obtainable. The memory experiments follow those described in the main text: we prepare the $[[20,2,6]]$ code using projective measurements in either the $\ket{\overline{00}}$ or $\ket{\overline{++}}$ states, perform $r=12$ rounds of syndrome extraction using the syndrome extraction circuit of Section~\ref{apx:syndrome_extraction}, and then destructively measure out the data qubits in the same basis in which they were prepared. From this final destructive measurement we can construct the syndromes and logical operators of a single type. We decode using Frontier and potentially perform postselection with $\gamma \le 4$. The logical error rate per round is then obtained as $\epsilon_L = 1 - (1 - p_L(r))^{1/r}$. As opposed to the main text, in this Appendix we report block error rate as opposed to a normalized per LQ error rate. Fig.~\ref{fig:memory_frontier}(a) displays $\epsilon_L$ as a function of the physical error rate, where we have averaged over the $X$ and the $Z$ bases. We fit the data to a power law and extract exponents that match reasonably well with the expected value of $\lfloor(d-1)/2\rfloor + 1$. Fig.\ref{fig:memory_frontier}(b) shows the postselection rate per round as a function of the physical error rate for varying $\gamma$. For $\gamma > 0$, the rejection rate per round exceeds the logical error rate per round, and so a long computation will be limited by accepted shots. However, for certain applications---in particular, those in which a small number of highly accurate shots are desirable---then the reduced logical error rate may justify the associated sampling overhead.

\begin{figure}
    \centering
    \includegraphics[width=0.95\linewidth]{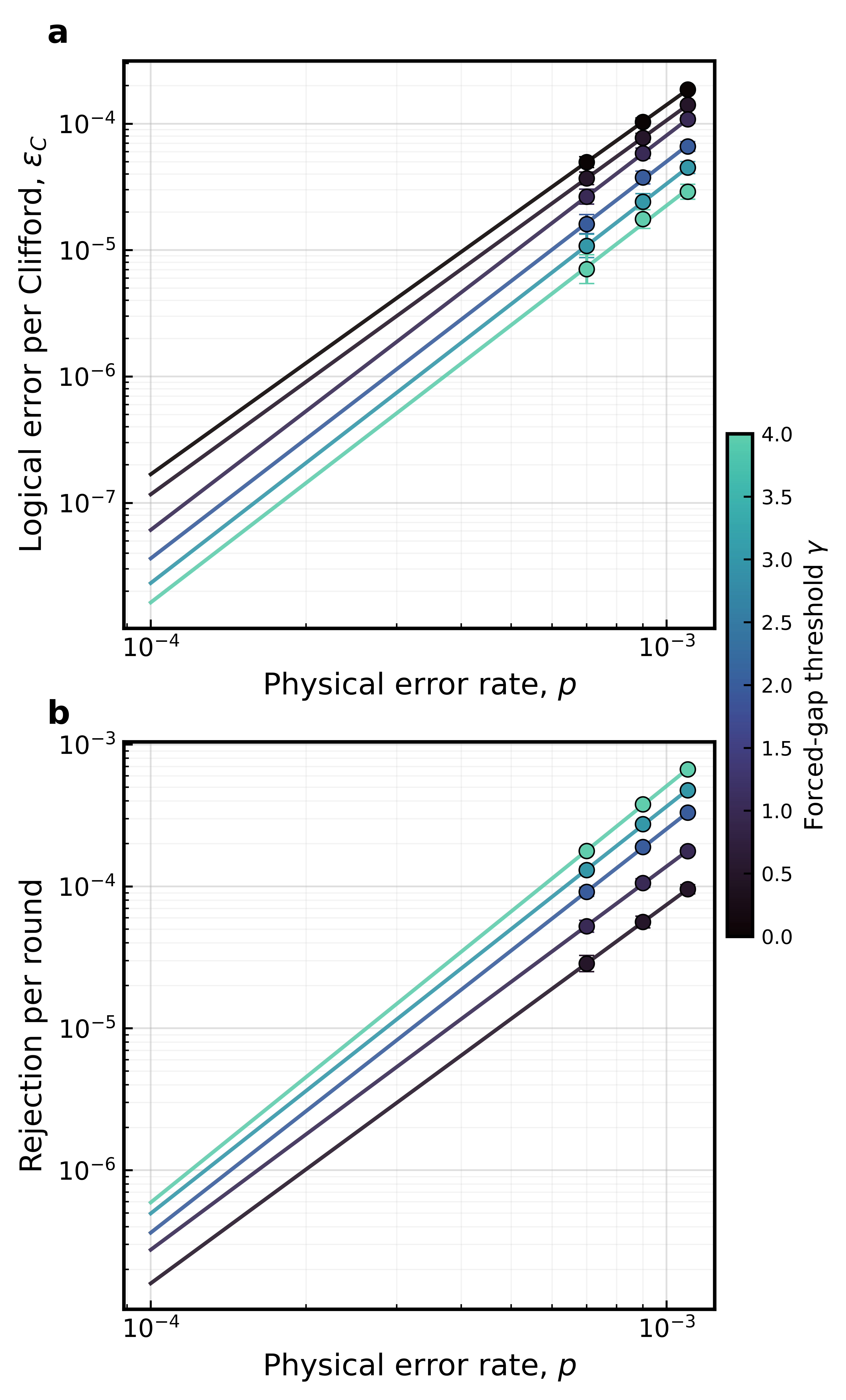}
    \caption{\textbf{Low physical error rate intrablock logic.} TQRB simulations of the \chelix\ code decoded using Frontier over the entire $m=6$ ($r=7$) Clifford sequence. \textbf{a.} Logical error rate per Clifford as a function of the physical error rate. \textbf{b.} Rejection rate per Clifford as a function of the physical error rate.}
    \label{fig:randomized_bencharking_sims}
\end{figure}

We additionally simulate randomized Clifford benchmarking to estimate the fidelity of performing logical intrablock Clifford operators. The simulations again follow the experiment run in the main text: we perform $m$ random logical two-qubit Clifford operators acting on a single \chelix\ codeblock. Having tracked the total accumulated Clifford operator, we calculate and apply a final inverse to return the code to the initial $\ket{\overline{00}}$ state. We again decode using Frontier over the entire circuit volume and optionally apply forced-gap postselection. The logical error rate per Clifford is calculated as above, but with $r = m+1$ `rounds' of logical gates. Fig.~\ref{fig:randomized_bencharking_sims}(a) reports the obtained logical error rates when extrapolated down to physical error rate $10^{-4}$. We see higher error rates for logic than for memory experiments which is expected due to a number of factors: each random Clifford contains on average 1.5 syndrome extraction rounds, the $\overline{S}_0 \overline{S}_1$ gate incorporates (flagged) in-block physical CZ gates which nonetheless present opportunities for errors to accumulate, and the decoding problem is more complex. We note that random intrablock Clifford operators are a worst-case workload for the \chelix\ code. In practice, clever physical-to-logical assignment can make it so most if not all intrablock Cliffords are automorphisms or transversal operations~\cite{LeBlond2026InPrep}. 

\begin{figure}
    \centering
    \includegraphics[width=0.9\linewidth]{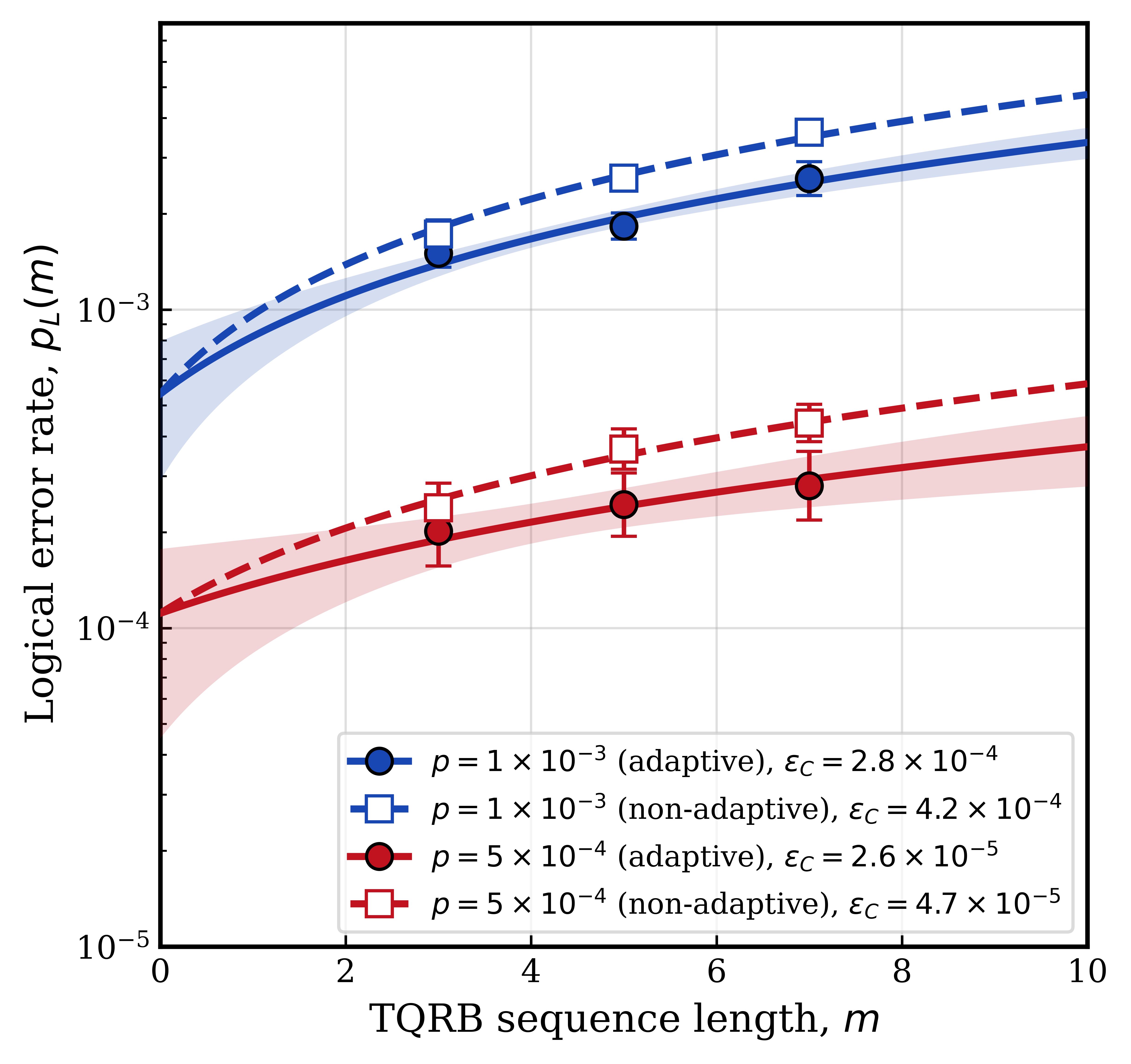}
    \caption{\textbf{Comparison of adaptive and nonadaptive syndrome extraction}. Logical error rate as a function of $m$, the TQRB sequence length. (Non)adaptive syndrome extraction is indicated by the (dashed) solid lines. We extract $\varepsilon_C$ by fitting to Eq.~\eqref{eq:sim_fit}. The SPAM offset term $\varepsilon_0$ is fixed between the adaptive and nonadaptive datasets.}
    \label{fig:logic_sims_adapt}
\end{figure}

The previous two simulations utilized nonadaptive syndrome extraction. To illustrate the advantages of adaptive syndrome extraction, we perform the same TQRB simulations but instead use the ASE scheme as described in Section~\ref{apx:syndrome_extraction}. We extract the logical error rate per Clifford $\varepsilon_C$ by fitting the individual $p_L(m)$ data points to 
\begin{equation}
    \label{eq:sim_fit}
    p_L^{(j)}(m) = \big(\varepsilon_0 - \frac{1}{2}\big)\big(1 - 2\varepsilon_C^{(j)}\big)^m + \frac{1}{2}.
\end{equation} 
We constrain the SPAM offset $\varepsilon_0$ to be identical between the datasets, as ASE should not affect this fit parameter. The results are shown in Fig.~\ref{fig:logic_sims_adapt}, where we see ASE outperforming its nonadaptive counterpart: we obtain a 1.5$\times$ reduction in $\varepsilon_C$ at physical error rate $p = 1 \times 10^{-3}$; this grows to a $1.8\times$ reduction at $p = 5\times 10^{-4}$. As the physical error rate continues to decrease, this performance gap will further increase. Additionally, the wall-clock savings increase as the physical error rate decreases: at $p = 1\times 10^{-3}$ ASE provides a 24\% shot-time reduction as compared to equivalent nonadaptive circuit. This increases to 30\% shorter shots at $p = 1\times 10^{-4}$. Even more significant shot time savings are possible in settings other than intrablock logic, e.g. memory or interblock transversal logic; in those settings there are no dangerous weight-2 errors to catch, and so the additional upper level generators are not required to be measured every round.

\begin{figure}[!t]
    \centering
    \includegraphics[width=0.85\linewidth]{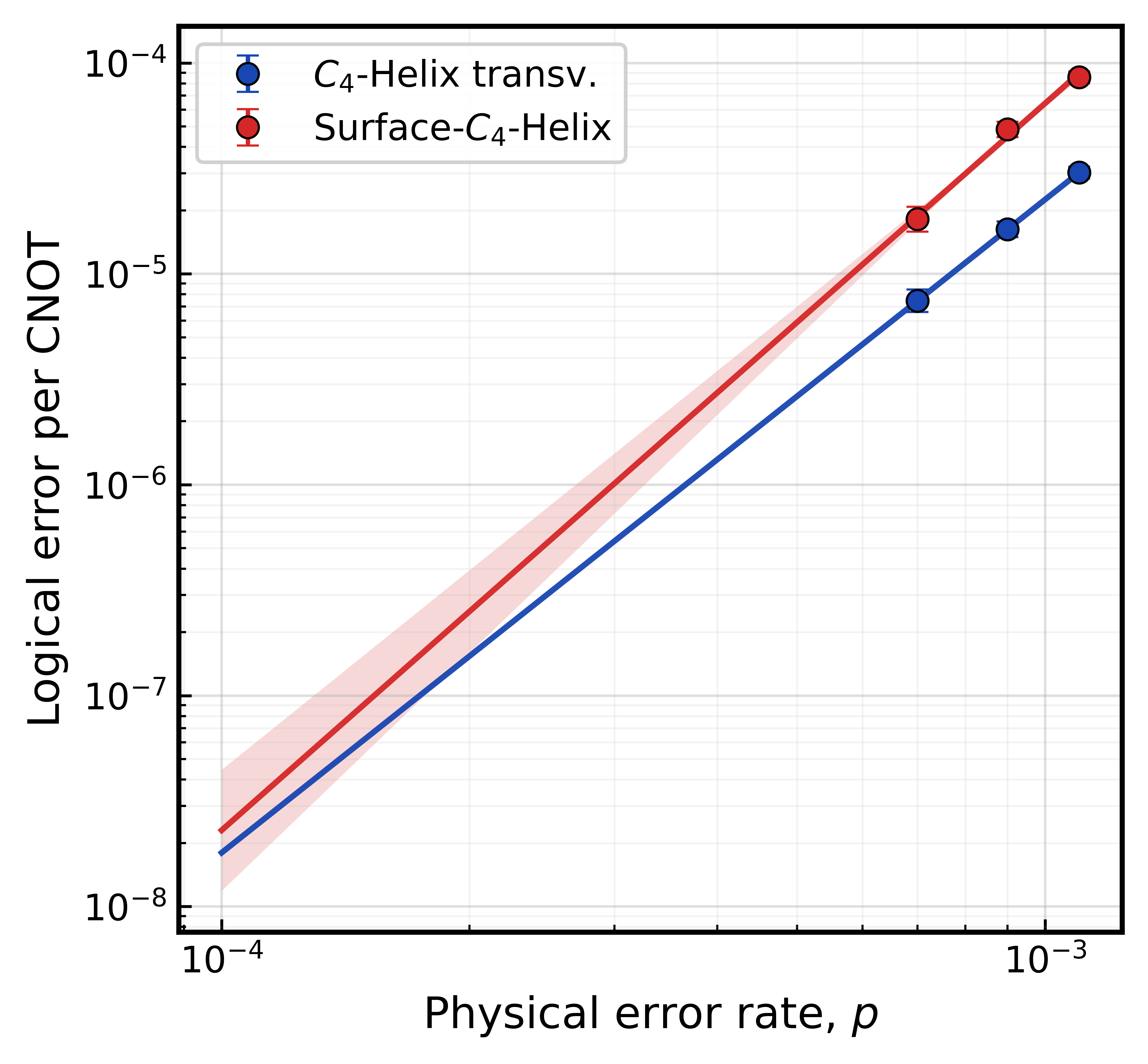}
    \caption{\textbf{Simulation of a homomorphic CNOT via chain maps.} A depth-2, distance-5 homomorphic CNOT from the distance-5 surface code to the first logical qubit of the \chelix\ code, compared with transversal CNOTs in the \chelix\ code. The full experiment consists of 6 transversal/homomorphic CNOT gates interleaved with QEC cycles on each code. Dashed lines indicate power-law fits. Shaded areas indicate 95\% Wilson confidence intervals.}
    \label{fig:two-blocks}
\end{figure}

We finally simulate inter-block and inter-code operations in the \chelix\ architecture. Since the \chelix\ code is a CSS code, it has a transversal CNOT gate in which 20 pairwise physical CNOT gates between two codeblocks implements two logical CNOT gates. We perform 6 transversal CNOT gates interleaved with QEC cycles on both codeblocks, and then we destructively measure in the basis in which we prepared them. As a comparison, we simulate the chain map described in Section~\ref{sec:ghz} which performs a logical CNOT between the distance-5 surface code and the first logical qubit of the \chelix\ code. We again perform 6 homomorphic CNOT gates interleaved with QEC cycles on both codes and destructively measure in the $Z$ basis before decoding over the entire circuit. The results are shown in Fig.~\ref{fig:two-blocks} which shows both logical gates obtaining a scaling exponent near 3, as expected. We make the interesting note that performing forced-gap postselection on this homomorphic CNOT circuit only yields an improvement in the logical error rate by about a factor of 2, whereas applying it in the memory, intrablock logic, and transversal CNOT settings is able to suppress logical error rates by up to two orders of magnitude. This weaker postselection benefit is consistent with the minimum failing-fault order of the distance-5 surface code: weight-3 faults can already produce competing logical classes, leaving less confidence separation for forced-gap postselection to exploit; and so logical error rates in this chain map are inherently limited by the surface code. 
